\pdfoutput=1
\RequirePackage{ifpdf}
\ifpdf 
\documentclass[pdftex]{sigma}
\else
\documentclass{sigma}
\fi

\numberwithin{equation}{section}

\begin{document}

\allowdisplaybreaks

\renewcommand{\PaperNumber}{075}

\FirstPageHeading

\ShortArticleName{Partner Symmetries, Group Foliation and ASD Ricci-Flat Metrics}

\ArticleName{Partner Symmetries, Group Foliation\\
and ASD Ricci-Flat Metrics without Killing Vectors}

\Author{Mikhail B.~SHEFTEL~$^\dag$ and Andrei A.~MALYKH~$^\ddag$}

\AuthorNameForHeading{M.B.~Sheftel and A.A.~Malykh}

\Address{$^\dag$~Department of Physics, Bo\u{g}azi\c{c}i University 34342 Bebek, Istanbul, Turkey}
\EmailD{\href{mailto:mikhail.sheftel@boun.edu.tr}{mikhail.sheftel@boun.edu.tr}}
\URLaddressD{\url{http://www.phys.boun.edu.tr/faculty_wp/mikhail_sheftel.html}}

\Address{$^\ddag$~Department of Numerical Modelling, Russian State Hydrometeorlogical University,\\
\hphantom{$^\ddag$}~98 Malookhtinsky Ave., 195196 St.~Petersburg, Russia} \EmailD{\href{mailto:andrei-malykh@mail.ru}{andrei-malykh@mail.ru}}

\ArticleDates{Received June 14, 2013, in f\/inal form November 19, 2013; Published online November 27, 2013}

\Abstract{We demonstrate how a~combination of our recently developed methods of partner symmetries,
symmetry reduction in group parameters and a~new version of the group foliation method can produce
noninvariant solutions of complex Monge--Amp\`ere equation (CMA) and provide a~lift from invariant
solutions of CMA satisfying Boyer--Finley equation to non-invariant ones.
Applying these methods, we obtain a~new noninvariant solution of CMA and the corresponding Ricci-f\/lat
anti-self-dual Einstein--K\"ahler metric with Euclidean signature without Killing vectors, together with
Riemannian curvature two-forms.
There are no singularities of the metric and curvature in a~bounded domain if we avoid very special choices
of arbitrary functions of a~single variable in our solution.
This metric does not describe gravitational instantons because the curvature is not concentrated in
a~bounded domain.}

\Keywords{Monge--Amp\`ere equation; Boyer--Finley equation; partner symmetries; symmetry reduction;
non-invariant solutions; group foliation; anti-self-dual gravity; Ricci-f\/lat metric}

\Classification{35Q75; 83C15}

\section{Introduction}

In his pioneer paper~\cite{Plebanski}, Pleba\~nski demonstrated that anti-self-dual (ASD) Ricci-f\/lat
metrics on four-dimensional complex manifolds are completely determined by a~single scalar potential which
satisf\/ies his f\/irst or second heavenly equation.
Such metrics are solutions to complex vacuum Einstein equations.
Real four-dimensional K\"ahler ASD metrics
\begin{gather}
ds^2=2\big(u_{1\bar1}dz^1d\bar{z}^1+u_{1\bar2}dz^1d\bar{z}^2+u_{2\bar1}dz^2d\bar{z}^1+u_{2\bar2}dz^2d\bar{z}^2\big)
\label{metric}
\end{gather}
that solve the vacuum Einstein equations with Euclidean (Riemannian) signature are governed by a~scalar
real-valued potential $u=u\big(z^1,z^2,\bar z^1,\bar z^2\big)$ which satisf\/ies complex Monge--Amp\`ere equation
(CMA)
\begin{gather}
u_{1\bar1}u_{2\bar2}-u_{1\bar2}u_{2\bar1}=1.
\label{cma}
\end{gather}
A modern justif\/ication for this conclusion one can f\/ind in the books by Mason and
Woodhouse~\cite{Mason/Woodhouse} and Dunajski~\cite{Dunajski}.
Here and henceforth, subscripts denote partial derivatives with respect to corresponding variables whereas
the bar means complex conjugation, e.g.\ $u_{1\bar 1} = \partial^2u/\partial z^1\partial\bar z^1$ and
suchlike.
The only exception is Section~\ref{sec-symred} where subscripts of generators $X$ are used to
designate dif\/ferent vector f\/ields.

Among ASD Ricci-f\/lat metrics, the most interesting ones are those that describe gravitational instantons
which asymptotically look like a~f\/lat space, so that their curvature is concentrated in a~f\/inite region
of a~Riemannian space-time (see~\cite{Dunajski} and references therein).
The most important gravitational instanton is $K3$ which geometrically is Kummer
surface~\cite{Atiyah/Hitchin/Singer}, for which an explicit form of the metric is still unknown while many
its properties and existence had been discovered and analyzed~\cite{Hitchin1974,Yau}.
A characteristic feature of the $K3$ instanton is that it does not admit any Killing vectors, that is, no
continuous symmetries which implies that the metric potential should be a~noninvariant solution of CMA
equation.
As opposed to the case of invariant solutions, for noninvariant solutions of CMA there should be no
symmetry reduction in the number of independent variables.
Since standard methods of Lie group analysis of PDEs provide only invariant solutions, which implies
symmetry reduction in the solutions and hence in the metric~\eqref{metric}, they cannot be applied for
obtaining noninvariant solutions to the CMA equation and ASD Ricci-f\/lat metrics without Killing vectors.
Thus, to obtain at least some pieces of $K3$ metric explicitly one needs a~technique for deriving
non-invariant solutions of multi-dimensional non-linear equations.

In the previous papers~\cite{Malykh/Nutku/Sheftel2003,Malykh/Nutku/Sheftel2004,Malykh/Sheftel2011} we have
developed methods for obtaining noninvariant solutions though still remaining in the symmetry framework.
We extended the Lax equations of Mason and Newman for CMA~\cite{Mason/Newman,Mason/Woodhouse} by
supplementing them with another pair of linear equations, so that CMA becomes an algebraic consequence of
these equations, whereas the original Lax pair generated only dif\/ferential consequences of CMA with no
distinction between elliptic, hyperbolic and homogeneous versions of CMA.
The equation, determining symmetry characte\-ris\-tics~\cite{Olver} for CMA, now appears as the integrability
condition of our linear equations.
Moreover, the symmetry condition has a~two-dimensional divergence form and therefore uniquely determines
locally a~potential function which turns out to be a~solution to the symmetry condition, that is, the
potential of a~symmetry is also a~symmetry, which is a~characteristic feature of a~more general class of
Monge--Amp\`ere equations~\cite{Sheftel/Malykh2009}.
We called this pair of symmetries, the original one and the corresponding potential, \textit{partner
symmetries} and applied them to generate noninvariant solutions of CMA and corresponding heavenly metrics
without Killing vectors.
We discovered that the equations connecting partner symmetries can be treated as an invariance condition
for solutions of CMA with respect to a~certain nonlocal symmetry constructed from partner symmetries and
nonlocal recursion operators~\cite{Malykh/Nutku/Sheftel2003}.
This was close but not identical to the idea of hidden symmetries by Dunajski and
Mason~\cite{Dunajski/Mason}.
The invariance of solutions under the nonlocal symmetry, unlike the invariance under a~local symmetry, does
not imply the symmetry reduction in the number of independent variables, so such solutions are noninvariant
in the usual sense.

The problem of f\/inding solutions of CMA by solving the equations for partner symmetries was facilitated
by the observation that the full system of such equations provides a~lift from invariant solutions of CMA
to its noninvariant solutions~\cite{Malykh/Nutku/Sheftel2007,Malykh/Sheftel2011,Sheftel/Malykh2008}.
As is shown in~\cite{Boyer/Finley}, any three dimensional reduction of CMA leads to either translationally
invariant solutions satisfying 3-dimensional Laplace equation or to rotationally invariant solutions
satisfying elliptic Boyer--Finley (BF) equation.
Earlier, in~\cite{Malykh/Sheftel2011}, we showed that there exists a~lift of translationally invariant
solutions of CMA to noninvariant solutions.
In~\cite{Malykh/Nutku/Sheftel2007} we obtained a~lift from hyperbolic version of Boyer--Finley equation to
noninvariant solutions of the hyperbolic CMA, with~1 replaced by~$-1$ on the right-hand side of~\eqref{cma}.
In this paper we will show that there also exist rotationally invariant solutions, related to solutions of
elliptic Boyer--Finley equation, which can be lifted to noninvariant solutions of the elliptic
CMA~\eqref{cma}.
Therefore, one may start with a~simpler problem of f\/inding translationally or rotationally invariant
solutions and then, using the partner symmetries, lift them to noninvariant solutions.

Another modif\/ication of the method of partner symmetries was, by using Lie equations, to introduce
explicitly symmetry group parameters (for commuting symmetries) as additional independent variables of the
problem instead of symmetry characteristics~\cite{Malykh/Sheftel2011}.
This trick allowed us to discover an integrability condition of the equations for partner symmetries.
Furthermore, this made possible using symmetry reductions in group parameters without symmetry reductions
in ``physical variables'' in order to simplify the equations to be solved without ending up with invariant
solutions.

In this paper, we apply to the full system of equations for partner symmetries of CMA another solution
tool, \textit{group foliation}~\cite{Martina/Sheftel/Winternitz,Nutku/Sheftel2001,Sheftel2002}, which we
used before for CMA and the Boyer--Finley equation~\cite{Boyer/Finley}.
The idea of the group foliation belongs to S.~Lie~\cite{Lie} and was developed by E.~Vessiot~\cite{Vessiot}
and a~modern presentation was given by L.V.
Ovsiannikov~\cite{Ovsiannikov}.
Original dif\/ferential equations are foliated with respect to a~chosen symmetry (sub)group into
\textit{automorphic} equations describing orbits of the symmetry group and \textit{resolving} equations
determining a~collection of the orbits.
The automorphic property of the f\/irst subsystem of equations means that any of its solutions can be
obtained from any other solution by some transformation of the chosen symmetry subgroup.
This property makes the automorphic system completely integrable if only one of its solutions can be
obtained.
Thus, the problem reduces to obtaining as many particular solutions of the resolving system as possible.
Each solution will f\/ix a~particular automorphic system and the corresponding orbit in the solution space
of original equations.

Applying this method to equations of partner symmetries, we f\/ind a~very large number of resolving
equations for which it is extremely dif\/f\/icult to f\/ind even a~single solution.
Therefore, we use our modif\/ication of the method which utilizes to a~greater degree \textit{operators of
invariant differentiation}.
They are def\/ined by the property that they commute with any prolongation of the symmetry generators of
the Lie group chosen for the foliation and hence they map dif\/ferential invariants again into
dif\/ferential invariants.
Our discovery was that the resolving system is equivalent to the commutator algebra of operators of
invariant dif\/ferentiation together with its Jacobi identities, so that we can replace the problem of
solving the resolving equations by the problem of solving commutator relations between the operators of
invariant dif\/ferentiation.
An example of such approach was given in our
papers~\cite{Martina/Sheftel/Winternitz,Nutku/Sheftel2001,Sheftel2002}.

In Section~\ref{sec-eqs}, we derive the extended system of six equations for partner symmetries of
CMA including their integrability condition and introduce symmetry group parameters as additional
independent variables.

In Section~\ref{sec-symred}, we determine all point symmetries of the extended system and perform
its symmetry reduction with respect to two group parameters.
This is not a~symmetry reduction with respect to ``physical'' variables and so it does not imply a~symmetry
reduction of solutions of CMA.
Our aim here is to prepare the ground for a~lift of some solutions to the elliptic version of the
Boyer--Finley (BF) equation to noninvariant solutions of CMA.
We determine all point symmetries of the reduced system.
To have BF in our system, we choose the symmetry of simultaneous rotations in complex $z^1$- and
$z^2$-planes for a~further reduction which picks up rotationally invariant solutions of CMA.
Being followed by a~Legendre transformation, this yields the BF equation.
Meanwhile, we also keep non-reduced transformed CMA equation obtained from the integrability condition of
the extended system, though in dif\/ferent variables involving symmetry group parameters.
If we f\/ind a~solution to BF which also satisf\/ies all other equations of the extended system, this would
mean a~lift from rotationally invariant solutions to noninvariant solutions of CMA.
Finally, we determine all point symmetries of the transformed extended system which we will need for the
group foliation.

In Section~\ref{sec-invdif}, we choose a~complex conjugate pair of the symmetry generators which
contain maximum number of arbitrary functions and thus generate a~maximal inf\/inite symmetry subgroup for
the group foliation.
We determine all f\/irst-order operators of invariant dif\/ferentiation and obtain a~set of all
dif\/ferential invariants up to the second order, inclusive.

In Section~\ref{sec-autres}, we perform the group foliation deriving the full set of automorphic and
resolving equations.
We derive also the commutator algebra of operators of invariant dif\/ferentiation.
There are too many resolving equations for a~straightforward search of its particular solutions.

In Section~\ref{sec-sol}, we f\/ind some solutions of the extended system by applying our strategy
of making ansatzes, that simplify the commutator algebra of operators of invariant dif\/ferentiation,
rather than trying to solve directly a~huge system of the resolving equations.
As a~by-product, we obtain new solutions of the Boyer--Finley equation.
We choose more generally looking solution for lifting it to a~solution of CMA.

In Section~\ref{sec-CMAsol}, we apply Legendre transformations of our solution to a~solution of CMA
equation with a~parameter-dependent right-hand side.
This solution depends on three arbitrary functions of a~single complex variable together with their complex
conjugates.
This is also a~simultaneous solution of the transformed Boyer--Finley equation in somewhat dif\/ferent
variables which is related to a~symmetry reduction of CMA and hence determines its invariant solutions.
Therefore, our construction provides a~lift from invariant to non-invariant solutions of CMA.

In Section~\ref{sec-metric}, we use our solution of the CMA equation to obtain K\"ahler metric of
Euclidean signature.
This metric is anti-self-dual and Ricci-f\/lat and have pole singularities in a~bounded domain only for
a~special choice of arbitrary functions in our solution.
By avoiding this special choice we obtain the metric without such singularities.
We have also computed Riemannian curvature two-forms which do not depend on two variables and hence the
curvature does not vanish outside of a~bounded domain in the space of all variables.
This means that our metric does not describe a~gravitational instanton.
Even though this simplest possible example of application of our approach does not produce an instanton
metric, we believe that more ref\/ined solutions to our extended system of equations for partner symmetries
and/or dif\/ferent chain of reductions will yield a~gravitational instanton metric with no Killing vectors.

In Section~\ref{sec-invsol}, we derive the invariance conditions for our solution under the
symmetries of the parameter-dependent CMA.
A detailed study of these conditions proves that generically (without special restrictions on arbitrary
functions in our solution to CMA) the solution is noninvariant and hence the corresponding metric has no
Killing vectors.

\section{Basic equations}
\label{sec-eqs}

In this section, we derive a~complete set of equations for partner symmetries to provide a~tool for
obtaining noninvariant solutions of CMA while remaining in the symmetry group
frame\-work~\mbox{\cite{Malykh/Nutku/Sheftel2003,Malykh/Nutku/Sheftel2004}}.
Here we introduce symmetry group parameters as additional independent variables in order to reserve
a~possibility of symmetry reduction in group parameters to simplify these equations without reduction in
the number of original independent variables in CMA and thus avoiding ending up with invariant solutions.
Using these additional variables also facilitates a~derivation of the integrability condition for our
equations, so that the extended set of equations augmented with this integrability condition is integrable
in the sense of Frobenius: all further integrability conditions are direct algebraic or dif\/ferential
consequences of already available equations.

For the complex Monge--Amp\`ere equation~\eqref{cma} the symmetry condition, that determines symmetry
characteristics $\varphi$ of~\eqref{cma},
\begin{gather}
u_{1\bar1}\varphi_{2\bar2}+u_{2\bar2}\varphi_{1\bar1}-u_{1\bar2}\varphi_{2\bar1}-u_{2\bar1}\varphi_{1\bar2}
=0
\label{symcond}
\end{gather}
on solutions of CMA.
Here the subscripts of $\varphi$ denote total derivatives with respect to corresponding independent
variables.
Equation~\eqref{symcond} can be set in the total divergence form
\begin{gather}
\label{div}
(u_{1\bar1}\varphi_2-u_{2\bar1}\varphi_1)_{\bar2}-(u_{1\bar2}\varphi_2-u_{2\bar2}\varphi_1)_{\bar1}=0.
\end{gather}
We assume that all functions that we operate with are smooth, that is, they have continuous derivatives of
any required order.
Then equation~\eqref{div} suggests \textit{local} existence of potential $\psi$ def\/ined by the equations
\begin{gather}
\psi_{\bar1}=u_{1\bar1}\varphi_2-u_{2\bar1}\varphi_1,
\qquad
\psi_{\bar2}=u_{1\bar2}\varphi_2-u_{2\bar2}\varphi_1
\label{12}
\end{gather}
in the sense that the condition~\eqref{div} becomes just the equality of mixed derivatives $(\psi_{\bar
1})_{\bar 2} = (\psi_{\bar 2})_{\bar 1}$, together with the complex conjugate equations
\begin{gather}
\bar\psi_1=u_{1\bar1}\bar\varphi_{\bar2}-u_{1\bar2}\bar\varphi_{\bar1},
\qquad
\bar\psi_2=u_{2\bar1}\bar\varphi_{\bar2}-u_{2\bar2}\bar\varphi_{\bar1}.
\label{b12}
\end{gather}
We do not discuss here the dif\/f\/icult problem of the global existence of potential $\psi$.

A straightforward check shows that the potential $\psi$ also satisf\/ies symmetry
condition~\eqref{symcond}, so that $\psi$ is also a~symmetry if $\varphi$ is a~symmetry and hence the
relations~\eqref{12} and~\eqref{b12} are recursion relations for symmetries (``partner symmetries'').
Transformation~\eqref{12} is algebraically invertible since its determinant equals one due to~\eqref{cma}.
Inverse transformation has the form
\begin{gather}
\varphi_1=u_{1\bar2}\psi_{\bar1}-u_{1\bar1}\psi_{\bar2},
\qquad
\varphi_2=u_{2\bar2}\psi_{\bar1}-u_{2\bar1}\psi_{\bar2}
\label{34}
\end{gather}
together with its complex conjugate
\begin{gather}
\bar\varphi_{\bar1}=u_{2\bar1}\bar\psi_1-u_{1\bar1}\bar\psi_2,
\qquad
\bar\varphi_{\bar2}=u_{2\bar2}\bar\psi_1-u_{1\bar2}\bar\psi_2.
\label{b34}
\end{gather}
For symmetries with characteristics $\varphi$, $\bar\varphi$, $\psi$ and $\bar\psi$, Lie equations read
\begin{gather}
\varphi=u_\tau,
\qquad
\bar\varphi=u_{\bar\tau},
\qquad
\psi=u_\sigma,
\qquad
\bar\psi=u_{\bar\sigma},
\label{Lie}
\end{gather}
where $\tau$, $\sigma$ together with their complex conjugates $\bar\tau$, $\bar\sigma$ are group parameters.
Simultaneous inclusion of several group parameters as additional independent variables implies the
commutativity conditions for corresponding symmetries in the form
\begin{gather*}
\varphi_{\bar\tau}=\bar\varphi_\tau,
\qquad
\psi_{\bar\sigma}=\bar\psi_\sigma,
\qquad
\varphi_\sigma=\psi_\tau,
\qquad
\varphi_{\bar\sigma}=\bar\psi_\tau
\end{gather*}
and complex conjugates to the last two equations.

We now use~\eqref{Lie} to replace symmetry characteristics by derivatives of $u$ with respect to group
parameters in equations~\eqref{12},~\eqref{34} and their complex conjugates~\eqref{b12},~\eqref{b34} with
the result
\begin{gather}
u_{\sigma\bar1}=u_{1\bar1}u_{\tau2}-u_{2\bar1}u_{\tau1},
\qquad
u_{\sigma\bar2}=u_{1\bar2}u_{\tau2}-u_{2\bar2}u_{\tau1},
\label{12a}
\\
u_{\tau1}=u_{1\bar2}u_{\sigma\bar1}-u_{1\bar1}u_{\sigma\bar2},
\qquad
u_{\tau2}=u_{2\bar2}u_{\sigma\bar1}-u_{2\bar1}u_{\sigma\bar2},
\label{34a}
\end{gather}
and the complex conjugate equations
\begin{gather}
u_{\bar\sigma1}=u_{1\bar1}u_{\bar\tau\bar2}-u_{\bar21}u_{\bar\tau\bar1},
\qquad
u_{\bar\sigma2}=u_{\bar12}u_{\bar\tau\bar2}-u_{2\bar2}u_{\bar\tau\bar1},
\label{b12a}
\\
u_{\bar\tau\bar1}=u_{\bar12}u_{\bar\sigma1}-u_{1\bar1}u_{\bar\sigma2},
\qquad
u_{\bar\tau\bar2}=u_{2\bar2}u_{\bar\sigma1}-u_{\bar21}u_{\bar\sigma2}.
\label{b34a}
\end{gather}
We note that four equations~\eqref{34a} and~\eqref{b34a} are algebraic consequences of other
equa\-tions~\eqref{12a}, \eqref{b12a} and CMA.
We note also that CMA itself follows as an algebraic consequence from equa\-tions~\eqref{12a}, \eqref{b12a}
and the f\/irst equation in~\eqref{b34a}.

To study integrability conditions of our system, we set the f\/irst equations in~\eqref{12a}
and~\eqref{b34a} in the form
\begin{gather}
(u_{1\bar1}u_2)_\tau=(u_\sigma+u_2u_{\tau1})_{\bar1},
\qquad
(u_{1\bar1}u_2)_{\bar\sigma}=(u_2u_{\bar\sigma1}-u_{\bar\tau})_{\bar1}.
\label{active}
\end{gather}
Equations~\eqref{active} constitute an active system since they have a~second-order nontrivial
integrability condition obtained by cross dif\/ferentiation of these equations with respect to $\bar\sigma$
and $\tau$ and further integration with respect to ${\bar z}^1$
\begin{gather*}
u_{\tau\bar\tau}+u_{\sigma\bar\sigma}+u_{\bar\sigma2}u_{\tau1}-u_{\bar\sigma1}u_{\tau2}=0,
\end{gather*}
where the ``constant'' of integration can be eliminated by a~redef\/inition of $u$.
To make this equation self-conjugate, we multiply it with an overall factor $u_{1\bar 1}$ and then
eliminate~$u_{1\bar 1}u_{\bar\sigma 2}$ and~$u_{1\bar 1}u_{\tau 2}$ in the last two terms of~\eqref{active}
using f\/irst equations in~\eqref{b34a} and~\eqref{12a}, respectively, with the f\/inal form of the
integrability condition
\begin{gather}
u_{1\bar1}(u_{\tau\bar\tau}+u_{\sigma\bar\sigma})-u_{\tau1}u_{\bar\tau\bar1}-u_{\sigma\bar1}u_{\bar\sigma1}
=0.
\label{sixeq}
\end{gather}
In a~similar way, we obtain the alternative form of the integrability condition
\begin{gather*}
u_{2\bar2}(u_{\tau\bar\tau}+u_{\sigma\bar\sigma})-u_{\tau2}u_{\bar\tau\bar2}-u_{\sigma\bar2}u_{\bar\sigma2}
=0.
\end{gather*}
We can choose~\eqref{12a},~\eqref{b12a},~\eqref{sixeq} and CMA for the set of algebraically independent
equations.
All other equations are linearly dependent on the chosen equations.
One could also check that there are no further independent second-order integrability conditions of our
system of six equations.

\section{Reduction of partner symmetries system for CMA}
\label{sec-symred}

Here we study point symmetries of the extended system for partner symmetries and use two symmetry
reductions with respect to group parameters to simplify this system.
The choice of the f\/irst reduction obeys the requirement for the reduced integrability condition~\eqref{8}
to be related to CMA~\eqref{ma} in new variables.
The second symmetry reduction of the extended system results in rotational reduction of the original
CMA~\eqref{cma} which yields equation~\eqref{cmarot} related to the Boyer--Finley equation~\eqref{e1} by
Legendre transformation~\eqref{legendre} combined with the dif\/ferential substitution $w=v_t$ and the
following integration of equations with respect to~$t$.
The transformed reduced integrability condition~\eqref{8a} becomes the Legendre transform of the
Monge--Amp\`ere equation~\eqref{modcma}.
In this way we arrive at the system which contains both CMA and Boyer--Finley equation (BF) thus providing
the possibility of a~lift from rotationally invariant solutions of CMA (related to solutions of BF) to
noninvariant solutions of CMA.
The Legendre transformation appears as a~necessary step because of the well-known relation between the
rotational reduction of CMA and BF equations (see, e.g.,~(4.10), (4.12) in~\cite{Boyer/Finley}).

We list the generators of all point symmetries of the extended system of six equations
CMA,~\eqref{12a},~\eqref{b12a} and~\eqref{sixeq}
\begin{gather}
X_1=\partial_\tau,
\qquad
\bar X_1=\partial_{\bar\tau},
\qquad
X_2=\partial_\sigma,
\qquad
\bar X_2=\partial_{\bar\sigma},
\qquad
X_3=\tau\partial_\tau+\sigma\partial_\sigma,
\nonumber
\\
\bar X_3=\bar\tau\partial_{\bar\tau}+\bar\sigma\partial_{\bar\sigma},
\qquad
X_4=z^2\partial_2-\bar{z}^2\partial_{\bar2}+\bar\tau\partial_{\bar\tau}
-\tau\partial_\tau+\sigma\partial_\sigma-\bar\sigma\partial_{\bar\sigma},
\nonumber
\\
X_5=\tau\partial_{\bar\sigma}-\sigma\partial_{\bar\tau},
\qquad
\bar X_5=\bar\tau\partial_\sigma-\bar\sigma\partial_\tau
\qquad
X_6=z^2\partial_2+\bar{z}^2\partial_{\bar2}+u\partial_u,
\nonumber
\\
X_a=a\big(z^1,z^2,\bar\tau,\sigma\big)\partial_u,
\qquad
X_{\bar a}=\bar a\big(\bar{z}^1,\bar{z}^2,\tau,\bar\sigma\big)\partial_u,
\label{symgen}
\\
X_c=c_{z^1}\partial_2-c_{z^2}\partial_1+(\tau c_\sigma-\bar\sigma c_{\bar\tau})\partial_u,
\nonumber
\\
X_{\bar c}=\bar c_{\bar{z}^1}\partial_{\bar2}-\bar c_{\bar{z}^2}\partial_{\bar1}
+(\bar\tau\bar c_{\bar\sigma}-\sigma\bar c_\tau)\partial_u,
\qquad
X_f=f(\tau,\sigma,\bar\tau,\bar\sigma)\partial_u,
\nonumber
\end{gather}
where $a$, $\bar a$, $c=c\big(z^1,z^2,\bar\tau,\sigma\big)$, $\bar c = \bar c\big(\bar{z}^1,\bar{z}^2,\tau,\bar\sigma\big)$
are arbitrary functions and $f(\tau,\sigma,\bar\tau,\bar\sigma)$ sa\-tisf\/ies the equation $f_{\tau\bar\tau}
+ f_{\sigma\bar\sigma} = 0$.
We note that obvious translational symmetry generators $\partial_1$, $\partial_{\bar 1}$, $\partial_2$ and
$\partial_{\bar 2}$ are particular cases of the generators $X_c$ and $X_{\bar c}$.
We have to emphasize that the subscripts of $X$ designate dif\/ferent vector f\/ields, contrary to our
previous convention that subscripts denote partial derivatives.

We specify two symmetries from~\eqref{symgen} for a~symmetry reduction of the extended system
\begin{gather}
X_I=\partial_\tau-\partial_1,
\qquad
\bar X_I=\partial_{\bar\tau}-\partial_{\bar1}.
\label{sym2}
\end{gather}
Solutions of CMA invariant with respect to symmetries~\eqref{sym2} are determined by the conditions
\begin{gather}
u_\tau=u_1,
\qquad
u_{\bar\tau}=u_{\bar1}.
\label{invcond}
\end{gather}
Using~\eqref{invcond}, we eliminate $u_\tau$ and $u_{\bar\tau}$ in all the
equations~\eqref{12a},~\eqref{b12a} and~\eqref{sixeq} to obtain
\begin{gather}
u_{\sigma\bar1}=u_{1\bar1}u_{12}-u_{2\bar1}u_{11},
\qquad
u_{\sigma\bar2}=u_{1\bar2}u_{12}-u_{2\bar2}u_{11},
\label{I_II}
\\
u_{\bar\sigma1}=u_{1\bar1}u_{\bar1\bar2}-u_{\bar21}u_{\bar1\bar1},
\qquad
u_{\bar\sigma2}=u_{\bar12}u_{\bar1\bar2}-u_{2\bar2}u_{\bar1\bar1},
\label{bI_II}
\\
u_{1\bar1}u_{\sigma\bar\sigma}-u_{1\bar\sigma}u_{\bar1\sigma}=u_{11}u_{\bar1\bar1}-u_{1\bar1}^2.
\label{8}
\end{gather}
We note that equation~\eqref{8} can be obtained by the Legendre transformation
\begin{gather*}
v=u-z^1u_1-\bar{z}^1u_{\bar1},
\qquad
p=-u_1,
\qquad
\bar p=-u_{\bar1}
\end{gather*}
of the CMA in new variables
\begin{gather}
\label{ma}
v_{p\bar p}v_{\sigma\bar\sigma}-v_{p\bar\sigma}v_{\sigma\bar p}=1.
\end{gather}

All point symmetry generators of the system of equations CMA,~\eqref{I_II},~\eqref{bI_II} and~\eqref{8} are
listed below
\begin{gather}
X_1=z^1\partial_1-\bar{z}^1\partial_{\bar1}-2\big(z^2\partial_2-\bar{z}^2\partial_{\bar2}\big),
\qquad
X_2=z^2\partial_2+\bar{z}^2\partial_{\bar2}+u\partial_u,
\nonumber
\\
X_3=a(\sigma)\partial_2+\frac{1}{2\lambda}\big(z^1\big)^2a'(\sigma)\partial_u,
\qquad
X_4=b(\bar\sigma)\partial_{\bar2}+\frac{\lambda}{2}\big(\bar{z}^1\big)^2b'(\bar\sigma)\partial_u,
\nonumber
\\
X_5=c'(\sigma)\big(z^1\partial_1-z^2\partial_2\big)+c(\sigma)\partial_\sigma-\frac{1}{2\lambda}
\big(z^1\big)^2z^2c''(\sigma)\partial_u,
\nonumber
\\
X_6=d'(\bar\sigma)\big(\bar{z}^1\partial_{\bar1}-\bar{z}^2\partial_{\bar2}\big)+d(\bar\sigma)\partial_{\bar\sigma}
-\frac{\lambda}{2}\big(\bar{z}^1\big)^2\bar{z}^2d''(\bar\sigma)\partial_u,
\nonumber
\\
X_7=-\lambda f_{z^2}\big(z^2,\sigma\big)\partial_1+z^1f_\sigma\big(z^2,\sigma\big)\partial_u,
\nonumber
\\
X_8=-\frac{1}{\lambda}g_{\bar{z}^2}\big(\bar{z}^2,\bar\sigma\big)\partial_{\bar1}+\bar{z}^1g_{\bar\sigma}\big(\bar{z}
^2,\bar\sigma\big)\partial_u,
\nonumber
\\
X_9=h\big(z^2,\sigma\big)\partial_u,
\qquad
X_{10}=k\big(\bar{z}^2,\bar\sigma\big)\partial_u.
\label{pointsym}
\end{gather}

To arrive at the Boyer--Finley equation, we need rotationally invariant solutions of
CMA~\cite{Boyer/Finley}.
Among the symmetries~\eqref{pointsym} of our extended system we can choose the symmetry of simultaneous
rotations in $z^1$ and $z^2$ complex planes, generated by $X_1$, which is convenient to combine with the
point transformation $z^2=e^p$, $\bar z^2=e^{\bar p}$.
The symmetry generator becomes
$X=- i X_1= 2i(\partial_p - \partial_{\bar p}) - i(z^1\partial_1 - \bar z^1\partial_{\bar 1})$, so that
new invariant variables are $\rho=p+\bar p$, $q=z^1e^{p/2}$, $\bar q=\bar z^1e^{\bar p/2}$ and
$u=u(\rho,q,\bar q,\sigma,\bar\sigma)$.

After this symmetry reduction the equations CMA,~\eqref{I_II},~\eqref{bI_II} and~\eqref{8} become
respectively
\begin{gather}
u_{q\bar q}u_{\rho\rho}-u_{\rho q}u_{\rho\bar q}=e^{\rho/2},
\label{cmarot}
\\
u_{\sigma\bar q}=u_{q\bar q}(u_{\rho q}+u_q/2)-u_{qq}u_{\rho\bar q},
\label{Ia}
\\
u_{\sigma\rho}=u_{\rho q}(u_{\rho q}+u_q/2)-u_{qq}u_{\rho\rho}
\label{IIa}
\end{gather}
together with complex conjugates of~\eqref{Ia} and~\eqref{IIa} and, f\/inally,
\begin{gather}
\label{8a}
u_{q\bar q}u_{\sigma\bar\sigma}-u_{\sigma\bar q}u_{\bar\sigma q}=e^{\rho/2}\big(u_{qq}u_{\bar q\bar q}
-u_{q\bar q}^2\big).
\end{gather}
We note that equation~\eqref{8a} has the form of the Legendre transform~\eqref{8} of the Monge--Amp\`ere
equation in variables~$q$, $\bar q$, $\sigma$, $\bar\sigma$ with~$\rho$ playing the role of a~parameter, similarly
to our remark after equation~\eqref{8}.
Parameter~$\rho$ can be scaled away by changing~$\sigma$,~$\bar\sigma$ to the new variables~$s$,~$\bar s$
def\/ined by $s = \sigma e^{\rho/4}$, $\bar s = \bar\sigma e^{\rho/4}$ when the equation~\eqref{8a} becomes
\begin{gather*}
u_{q\bar q}u_{s\bar s}-u_{s\bar q}u_{\bar sq}=u_{qq}u_{\bar q\bar q}-u_{q\bar q}^2,
\end{gather*}
which is exactly the Monge--Amp\`ere equation
\begin{gather}
\label{modcma}
v_{p\bar p}v_{s\bar s}-v_{p\bar s}v_{s\bar p}=1
\end{gather}
after the Legendre transformation
\begin{gather}
\label{legmod}
v=u-qu_q-\bar qu_{\bar q},
\qquad
p=-u_q,
\qquad
\bar p=-u_{\bar q}.
\end{gather}
Therefore, our system contains also transformed Monge--Amp\`ere equation~\eqref{modcma}, though in
dif\/ferent variables, in the form~\eqref{8a}.

Our aim is to transform the reduced CMA to the Boyer--Finley equation.
On the way to it, we apply one-dimensional Legendre transformation
\begin{gather}
\label{legendre}
t=u_\rho,
\qquad
w=u-\rho u_\rho,
\qquad
\rho=-w_t,
\qquad
u=w-tw_t.
\end{gather}
CMA equation~\eqref{cmarot} becomes
\begin{gather*}
w_{q\bar q}=-w_{tt}e^{-w_t/2}.
\end{gather*}
Equations~\eqref{Ia} and~\eqref{IIa} take the form
\begin{gather}
\label{I_IIb}
w_{\sigma\bar q}=\frac{1}{2}w_qw_{q\bar q}+w_{tq}e^{-w_t/2},
\qquad
w_{t\sigma}=\frac{1}{2}w_qw_{tq}-w_{qq}
\end{gather}
and complex conjugate equations.
The image of equation~\eqref{8a} with the use of other equations becomes
\begin{gather}
\label{8b}
w_{\sigma\bar\sigma}=w_{tt}e^{-w_t}+\frac{1}{2}\left(w_{\bar q}w_{tq}+w_qw_{t\bar q}-\frac{1}{2}
w_qw_{\bar q}w_{tt}\right)e^{-w_t/2}.
\end{gather}

The f\/inal step is to set $w = v_t$, which makes all the equations to be total derivatives with respect to
$t$, and then integrate the equations with respect to $t$.
To simplify the notation, we also change $\sigma$ to $z$ and $\bar\sigma$ to $\bar z$.
The reduced CMA takes the form of the Boyer--Finley equation
\begin{gather}
\label{e1}
v_{q\bar{q}}=2e^{-v_{tt}/2}.
\end{gather}
Equations~\eqref{I_IIb} together with their complex conjugates read
\begin{gather}
v_{qq}=-v_{tz}+\frac{1}{4}v_{tq}^2,
\label{e2}
\\
v_{\bar{q}\bar{q}}=-v_{t\bar{z}}+\frac{1}{4}v_{t\bar{q}}^2,
\label{be2}
\\
v_{\bar{q}z}=v_{tq}e^{-v_{tt}/2},
\label{e3}
\\
v_{q\bar{z}}=v_{t\bar{q}}e^{-v_{tt}/2},
\label{be3}
\end{gather}
and the equation~\eqref{8b} becomes
\begin{gather}
\label{e4}
v_{z\bar{z}}=-e^{-v_{tt}}+\frac{1}{2}v_{tq}v_{t\bar{q}}e^{-v_{tt}/2}.
\end{gather}
According \looseness=-1 to our remark after equation~\eqref{legmod}, our f\/inal system contains besides Boyer--Finley
equation~\eqref{e1} also transformed Monge--Amp\`ere equation, which is a~consequence of this system,
though its explicit form, being a~bit lengthy, is not presented here.
Therefore, we have hopes that f\/inding some noninvariant solutions of the Boyer--Finley equation, we will
be able to lift them to noninvariant solutions of the complex Monge--Amp\`ere equation, this being our main
goal.

Point symmetries generators of the above system are
\begin{gather}
X_1=\partial_t,
\qquad
X_2=q\partial_q+\bar{q}\partial_{\bar{q}}+2t\partial_t+\big(4v-2t^2\big)\partial_v,\qquad X_3=qa(z)\partial_v,\qquad X_4=\bar{X}_3,
\nonumber\\
X_5=b(z)\partial_v,
\qquad
X_6=\bar{X}_5,
\qquad
X_7=\big(tc(z)-q^2c'(z)/2\big)\partial_v,
\qquad
X_8=\bar{X}_7,
\nonumber\\
X_9=d(z)\partial_q+\bigl\{q^3d''(z)/3-2qtd'(z)\bigr\}\partial_v,
\qquad
X_{10}=\bar{X}_9,\label{s10_12}
\\
X_{11}=\frac{1}{2}f'(z)q\partial_q+f(z)\partial_z+\bigl\{q^4f'''(z)/24-tq^2f''(z)/2+t^2f'(z)/2\bigr\}
\partial_v,
\qquad
X_{12}=\bar{X}_{11},\nonumber
\end{gather}
where the bars mean complex conjugation.
All functions are arbitrary and smooth.
No nontrivial contact symmetries exist.

\section{Operators of invariant dif\/ferentiation\\ and second-order
dif\/ferential invariants}
\label{sec-invdif}

For the \textit{group foliation}~\cite{Martina/Sheftel/Winternitz,Nutku/Sheftel2001,Sheftel2002} of the
system~\eqref{e1}--\eqref{e4} we choose the inf\/inite dimensional Lie subgroup generated by $X_{11}$ and
$\bar{X}_{11}$ since it contains maximum number of arbitrary functions and hence maximum number of
constraints for dif\/ferential invariants with respect to this subgroup.

In group foliation, an important role is played by \textit{operators of invariant differentiation} which,
by def\/inition, commute with any prolongation of symmetry generators $X_{11}$ and $\bar{X}_{11}$.
The number of independent operators of invariant dif\/ferentiation is the same as the number of independent
variables, that is, f\/ive in our case.
The equations determining such operators one can f\/ind in Ovsiannikov's book~\cite{Ovsiannikov} or in our
paper~\cite{Martina/Sheftel/Winternitz}.
Here we present only the result for solving these equations which f\/ixes the following form of operators
of invariant dif\/ferentiation:
\begin{gather}
\delta=D_t,
\qquad
\Delta^q=qD_q,
\qquad
\bar\Delta^q=\bar qD_{\bar q},
\nonumber
\\
\Delta^z=q^2(2D_z-v_{tq}D_q),
\qquad
\bar\Delta^z={\bar q}^2(2D_{\bar z}-v_{t\bar q}D_{\bar q}),
\label{invdif}
\end{gather}
where $D$ denotes total derivative with respect to its letter subscript.
Operators~\eqref{invdif}, when acting on an invariant, generate again a~(dif\/ferential) invariant,
increasing its order by one unit.
Invariant dif\/ferentiations may also generate dif\/ferential invariants even when acting on a~noninvariant
quantities (`pre-invariants').
A~\textit{basis of differential invariants} is formed by invariants such that invariant dif\/ferentiations
can generate any invariant of an arbitrary order by repeated applications to basis invariants.

A zeroth-order invariant is~$t$.
A single f\/irst-order invariant is
\begin{gather*}
\omega_1=qv_q+\bar qv_{\bar q}+2tv_t-4v.
\end{gather*}
The complete set of second-order independent dif\/ferential invariants consists of 12 invariants.
Indeed, the dimension of the $N$th prolongation space, where $N$ is the order of the prolongation is $\nu_N
= n + m\frac{(N+n)!}{N!n!}$, where $n$ and $m$ are the numbers of independent and dependent variables,
respectively, and $N = 1,2,3,\dots$.
In our case $n=5$, $m=1$, so $\nu_N = 5 + \frac{(N+5)!}{N!5!}$, which for $N=2$ yields $\nu_2=26$.
The dimension of orbits $r_N$ is the rank of the system of $N$th prolongations of generators $X_{11}$ and
$\bar{X}_{11}$, which is equal to the number of arbitrary functions of $z$, $\bar z$ that they contain.
For $N=2$, we have arbitrary functions in the second prolongation of our two generators
$f^{\prime\prime\prime\prime}(z)$, $f^{\prime\prime\prime\prime\prime}(z)$, $\bar f^{\prime\prime\prime\prime}(\bar z)$, $\bar f^{\prime\prime\prime\prime\prime}(\bar z)$ in addition to those which appear
in the last line of~\eqref{s10_12}, that is, $r_2 = 12$.
The dimension of the space of invariants is $\dim \mathbb{Z}_N = \nu_N - r_N$ and for $N=2$ this
becomes $\dim \mathbb{Z}_2 = \nu_2 - r_2 = 14$.
Therefore, in addition to the two invariants of zeroth and f\/irst-order we must obtain 12 independent
second-order invariants.
All of them are listed below:
\begin{gather}
\omega_2=q\bar qe^{-v_{tt}/2},
\label{om2}
\\
\omega_3=\Delta^q(\omega_1)=q(qv_{qq}+\bar qv_{q\bar q}+2tv_{tq}-3v_q),
\label{om3}
\\
\bar\omega_3=\bar\Delta^q(\omega_1)=\bar q(\bar qv_{\bar q\bar q}+qv_{q\bar q}+2tv_{t\bar q}-3v_{\bar q}),
\label{bom3}
\\
\omega_4=\delta(\omega_1)=qv_{tq}+\bar qv_{t\bar q}+2tv_{tt}-2v_t,
\label{om4}
\\
\omega_5=q\bar q v_{q\bar q}=\bar\Delta^q(qv_q)=\Delta^q(\bar qv_{\bar q}),
\label{om5}
\\
\omega_6=q^2\bar q(2v_{\bar qz}-v_{q\bar q}v_{tq}),
\qquad
\bar\omega_6=q{\bar q}^2(2v_{q\bar z}-v_{q\bar q}v_{t\bar q}),
\label{om6}
\\
\omega_7=q^2\left(v_{tz}+v_{qq}-v_{tq}^2/4\right),
\qquad
\bar\omega_7={\bar q}^2\left(v_{t\bar z}+v_{\bar q\bar q}-v_{t\bar q}^2/4\right),
\label{om7}
\\
\omega_8=q^3{\bar q}^3(v_{q\bar q}v_{z\bar z}-v_{q\bar z}v_{\bar qz}),
\label{om8}
\\
\omega_9=\Delta^z(\omega_1)=q^2\big\{4tv_{tz}+2qv_{qz}+2\bar qv_{\bar qz}-8v_z
\nonumber
\\
\hphantom{\omega_9=\Delta^z(\omega_1)=}{}-v_{tq}(qv_{qq}+\bar qv_{q\bar q}+2tv_{tq}-3v_q)\big\},
\label{om9}
\\
\bar\omega_9=\bar\Delta^z(\omega_1)={\bar q}^2\big\{4tv_{t\bar z}+2\bar qv_{\bar q\bar z}+2qv_{q\bar z}
-8v_{\bar z}
\nonumber
\\
\hphantom{\bar\omega_9=\bar\Delta^z(\omega_1)=}{}-v_{t\bar q}(\bar qv_{\bar q\bar q}+qv_{q\bar q}+2tv_{t\bar q}-3v_{\bar q})\big\}.
\label{bom9}
\end{gather}

\section{Automorphic and resolving equations}
\label{sec-autres}

Here we f\/ix the form of automorphic and resolving equations using only invariant variables.
We choose a~set of 5 independent invariant variables (same number as in the original
system~\eqref{e1}--\eqref{e4}) and three remaining dif\/ferential invariants are considered as new
invariant unknowns in the automorphic system~\eqref{autom} (see explanation below~\eqref{inveqs}).
Integrability conditions of these equations together with the original system yield resolving equations for
the unknown functions $F$, $G$ and $\bar G$ in~\eqref{autom}.
This task is much simplif\/ied by applying our modif\/ication of the method which uses commutator algebra
of the operators of invariant dif\/ferentiation together with its Jacobi
identi\-ties~\mbox{\cite{Martina/Sheftel/Winternitz,Nutku/Sheftel2001,Sheftel2002}}.
For any solution of resolving equations for $F$, $G$ and $\bar G$ the system~\eqref{autom} possesses the
automorphic property: any solution can be obtained from any other solution by a~symmetry group
transformation generated by~$X_{11}$ and~$\bar{X}_{11}$.

All our equations~\eqref{e1}--\eqref{e4} can be expressed solely in terms of dif\/ferential invariants as
follows:
\begin{gather}
\omega_5=2\omega_2
\quad
\eqref{e1},
\qquad
\omega_7=0
\quad
\eqref{e2},
\qquad
\bar\omega_7=0
\quad
\eqref{be2},
\qquad
\omega_6=0
\quad
\eqref{e3},
\nonumber
\\
\bar\omega_6=0
\quad
\eqref{be3},
\qquad
\omega_8=-2\omega_2^3
\quad
\eqref{e4}.
\label{inveqs}
\end{gather}
Hence out of the twelve second-order invariants~\eqref{om2}--\eqref{bom9} there are only six independent
ones.
Therefore, for the second--order of prolongation we have only eight independent invariants together with $t$
and $\omega_1$ on the solution manifold of our system of six equations.
For the group foliation, we should separate them into two groups: independent and dependent invariant
va\-riab\-les.
In order not to loose any solutions of our original equations, the number of independent invariant
variables should be f\/ive, the same as in the original equations.
We choose them to be $(t,\omega_1,\omega_2,\omega_3,\bar\omega_3)$.
Thus, only three remaining invariants should be chosen as new invariant unknown functions of the f\/ive
independent invariant variables
\begin{gather}
\omega_4=F(t,\omega_1,\omega_2,\omega_3,\bar\omega_3)=\delta(\omega_1),
\qquad
\omega_9=G(t,\omega_1,\omega_2,\omega_3,\bar\omega_3)=\Delta^z(\omega_1),
\nonumber
\\
\bar\omega_9=\bar G(t,\omega_1,\omega_2,\omega_3,\bar\omega_3)=\bar\Delta^z(\omega_1),
\label{autom}
\end{gather}
which is the general form of the automorphic system.
Some part of resolving equations we obtain from \textit{inner integrability conditions} for these three
equations together with the equations that follow from the def\/initions of $\omega_3$, $\bar\omega_3$
\begin{gather}
\label{om3def}
\omega_3=\Delta^q(\omega_1),
\qquad
\bar\omega_3=\bar\Delta^q(\omega_1).
\end{gather}
These conditions are obtained by \textit{invariant cross differentiation} of equations~\eqref{autom}
and~\eqref{om3def} where we will use the \textit{commutator algebra of operators of invariant
differentiation}
\begin{gather}
[\delta,\Delta^q]=[\delta,\bar\Delta^q]=0,
\qquad
[\delta,\Delta^z]=-\omega_{tz}\Delta^q,
\qquad
[\delta,\bar\Delta^z]=-\bar\omega_{tz}\bar\Delta^q,
\nonumber
\\
[\Delta^q,\Delta^z]=2\Delta^z-\omega_{qz}\Delta^q,
\qquad
[\bar\Delta^q,\bar\Delta^z]=2\bar\Delta^z-\bar\omega_{qz}\bar\Delta^q,
\nonumber
\\
[\Delta^q,\bar\Delta^q]=0,
\qquad
[\Delta^q,\bar\Delta^z]=\omega_2\omega_{tt}\bar\Delta^q,
\qquad
[\bar\Delta^q,\Delta^z]=\omega_2\omega_{tt}\Delta^q,
\nonumber
\\
[\Delta^z,\bar\Delta^z]=2\omega_2\big(\bar\omega_{tz}\Delta^q-\omega_{tz}\bar\Delta^q\big).\label{dD}
\end{gather}
Here coef\/f\/icients are the following third-order invariants
\begin{gather}
\omega_{tt}=v_{ttt}=\delta(v_{tt}),
\qquad
\omega_{tz}=qv_{ttq}=\Delta^q(v_{tt}),
\qquad
\bar\omega_{tz}=\bar qv_{tt\bar q}=\bar\Delta^q(v_{tt}),
\label{omtz}
\\
\omega_{qz}=q^2v_{tqq}-qv_{tq}=\Delta^q(qv_{tq})-2qv_{tq}
\nonumber
\\
\hphantom{\omega_{qz}}{} =-\big\{q^2(v_{ttz}-v_{tq}v_{ttq}/2)+qv_{tq}\big\}=-\frac{1}{2}\Delta^z(v_{tt})-qv_{tq},
\label{omqz}
\\
\bar\omega_{qz}=\bar q^2v_{t\bar q\bar q}-\bar qv_{t\bar q}=\bar\Delta^q(\bar qv_{t\bar q}
)-2\bar qv_{t\bar q}
\nonumber
\\
\hphantom{\bar\omega_{qz}}{} =-\big\{\bar q^2(v_{tt\bar z}-v_{t\bar q}v_{tt\bar q}/2)+\bar qv_{t\bar q}\big\}=-\frac{1}{2}\bar\Delta^z(v_{tt}
)-\bar qv_{t\bar q},
\label{ombqz}
\end{gather}
where the alternative expressions for $\omega_{qz}$ and $\bar\omega_{qz}$ in the second lines
of~\eqref{omqz} and~\eqref{ombqz} are obtained by using equations~\eqref{e2} and~\eqref{be2}, respectively.

In the integrability conditions of equations~\eqref{autom} and~\eqref{om3def} new third-order invariants
appear which are obtained by the action of operators of invariant dif\/ferentiation on second-order
invariants $\omega_2$, $\omega_3$, $\bar\omega_3$.
We introduce for them the following notation
\begin{gather}
\label{omij}
\omega_{it}=\delta(\omega_i),\!
\qquad
\omega_{iq}=\Delta^q(\omega_i),\!
\qquad
\omega_{i\bar q}=\bar\Delta^q(\omega_i),\!
\qquad
\omega_{iz}=\Delta^z(\omega_i),\!
\qquad
\omega_{i\bar z}=\bar\Delta^z(\omega_i),\!\!
\end{gather}
where $i = 2,3$, together with complex conjugates to the equations at $i=3$.

Equations~\eqref{omij} determine projections of operators of invariant dif\/ferentiation on the space of
invariants
\begin{gather*}
\delta=\partial_t+F\partial_{\omega_1}+\omega_{2t}\partial_{\omega_2}+\omega_{3t}\partial_{\omega_3}
+\bar\omega_{3t}\partial_{\bar\omega_3},
\\
\Delta^q=\omega_3\partial_{\omega_1}+\omega_{2q}\partial_{\omega_2}+\omega_{3q}\partial_{\omega_3}
+\bar\omega_{3\bar q}\partial_{\bar\omega_3},
\\
\bar\Delta^q=\bar\omega_3\partial_{\omega_1}+\bar\omega_{2q}\partial_{\omega_2}+\omega_{3\bar q}
\partial_{\omega_3}+\bar\omega_{3q}\partial_{\bar\omega_3},
\\
\Delta^z=G\partial_{\omega_1}+\omega_{2z}\partial_{\omega_2}+\omega_{3z}\partial_{\omega_3}
+\bar\omega_{3\bar z}\partial_{\bar\omega_3},
\\
\bar\Delta^z=\bar G\partial_{\omega_1}+\bar\omega_{2z}\partial_{\omega_2}+\omega_{3\bar z}
\partial_{\omega_3}+\bar\omega_{3z}\partial_{\bar\omega_3}.
\end{gather*}

Then invariant integrability conditions for equations~\eqref{autom} and~\eqref{om3def} can be obtained by
applying commutator relations between operators of invariant dif\/ferentiation~\eqref{dD} to the f\/irst-order invariant $\omega_1${\samepage
\begin{gather}
\Delta^q(F)=\omega_{3t},
\qquad
\bar\Delta^q(F)=\bar\omega_{3t},
\qquad
\Delta^z(F)=\delta(G)+\omega_3\omega_{tz},
\qquad
\bar\Delta^z(F)=\delta(\bar G)+\bar\omega_3\bar\omega_{tz},
\nonumber
\\
\Delta^q(G)=2G+\omega_{3z}-\omega_3\omega_{qz},
\qquad
\bar\Delta^q(\bar G)=2\bar G+\bar\omega_{3z}-\bar\omega_3\bar\omega_{qz},\qquad \bar\Delta^q(G)=\bar\omega_{3\bar z}-\omega_3\omega_{q\bar z},
\nonumber
\\
\Delta^q(\bar G)=\omega_{3\bar z}-\bar\omega_3\omega_{q\bar z},
\qquad
\bar\Delta^z(G)=\Delta^z(\bar G)+2\omega_2(\bar\omega_3\omega_{tz}-\omega_3\bar\omega_{tz})
\label{DFG}
\end{gather}
with the reality conditions $\bar\omega_{3\bar q} = \omega_{3\bar q}$ and $\bar\omega_{q\bar z} =
\omega_{q\bar z}$.}

These equations have already the form of resolving equations if we consider here third-order invariants
def\/ined in~\eqref{omij} as auxiliary unknowns which are functions of
$t$, $\omega_1$, $\omega_2$, $\omega_3$, $\bar\omega_3$.
Then we must add new resolving equations following from their def\/initions~\eqref{omij}.
Some of them we obtain by applying commutator relations~\eqref{dD} between invariant dif\/ferentiations to
independent second-order invariant variables $\omega_2$, $\omega_3$, $\bar\omega_3$ as follows
\begin{gather}
\Delta^q(\omega_{it})=\delta(\omega_{iq}),
\qquad
\bar\Delta^q(\omega_{it})=\delta(\bar\omega_{iq}),
\qquad
\bar\Delta^q(\omega_{iq})=\Delta^q(\bar\omega_{iq}),
\nonumber
\\
\Delta^z(\omega_{it})=\delta(\omega_{iz})-\omega_{tz}\omega_{iq},
\qquad
\bar\Delta^z(\bar\omega_{it})=\delta(\bar\omega_{iz})-\bar\omega_{tz}\bar\omega_{iq},
\nonumber
\\
\Delta^q(\omega_{iz})=\Delta^z(\omega_{iq})+2\omega_{iz}-\omega_{qz}\omega_{iq},
\qquad
\bar\Delta^q(\bar\omega_{iz})=\bar\Delta^z(\bar\omega_{iq})+2\bar\omega_{iz}-\bar\omega_{qz}\bar\omega_{iq},
\nonumber
\\
\Delta^q(\bar\omega_{iz})=\bar\Delta^z(\omega_{iq})-\omega_{q\bar z}\bar\omega_{iq},
\qquad
\bar\Delta^q(\omega_{iz})=\Delta^z(\bar\omega_{iq})-\omega_{q\bar z}\omega_{iq},
\nonumber
\\
\Delta^z(\bar\omega_{iz})=\bar\Delta^z(\omega_{iz})+2\omega_2(\bar\omega_{tz}\omega_{2q}-\omega_{tz}
\bar\omega_{2q}),
\label{Dom}
\end{gather}
where $i=2,3$, plus complex conjugate equations at $i=3$.
Here $\bar\omega_{2t} = \omega_{2t}$.

To obtain further resolving equations, we consider Jacobi identities between triples of ope\-rators of
invariant dif\/ferentiation using commutator relations~\eqref{dD}
\begin{gather}
\Delta^q(\omega_{tz})=\delta(\omega_{qz})+2\omega_{tz},\qquad \bar\Delta^q(\bar\omega_{tz}
)=\delta(\bar\omega_{qz})+2\bar\omega_{tz},
\nonumber
\\
\bar\Delta^q(\omega_{tz})=\Delta^q(\bar\omega_{tz})=-\omega_2\delta(\omega_{tt})-\omega_{2t}\omega_{tt},
\nonumber
\\
\Delta^z(\bar\omega_{tz})=2\omega_2\delta(\omega_{tz})+\omega_{tz}(2\omega_{2t}-\omega_2\omega_{tt}),
\nonumber
\\
\bar\Delta^z(\omega_{tz})=2\omega_2\delta(\bar\omega_{tz})+\bar\omega_{tz}(2\omega_{2t}-\omega_2\omega_{tt}),
\nonumber
\\
\Delta^q(\bar\omega_{qz})=-\omega_2\bar\Delta^q(\omega_{tt})+\omega_{tt}(2\omega_2-\bar\omega_{2q}),
\nonumber
\\
\bar\Delta^q(\omega_{qz})=-\omega_2\Delta^q(\omega_{tt})+\omega_{tt}(2\omega_2-\omega_{2q}),
\nonumber
\\
\Delta^z(\bar\omega_{qz})=2\omega_2\bar\Delta^q(\omega_{tz})-2\omega_{tz}(2\omega_2-\bar\omega_{2q}
)+\omega_2^2\omega_{tt}^2,
\nonumber
\\
\bar\Delta^z(\omega_{qz})=2\omega_2\Delta^q(\bar\omega_{tz})-2\bar\omega_{tz}(2\omega_2-\omega_{2q}
)+\omega_2^2\omega_{tt}^2,
\nonumber
\\
\omega_2\big\{\Delta^z(\omega_{tt})+2\Delta^q(\omega_{tz})\big\}=-\omega_{tt}(\omega_{2z}+\omega_2\omega_{qz}
)-2\omega_{tz}(2\omega_2+\omega_{2q}),
\nonumber
\\
\omega_2\big\{\bar\Delta^z(\omega_{tt})+2\bar\Delta^q(\bar\omega_{tz})\big\}=-\omega_{tt}(\bar\omega_{2z}
+\omega_2\bar\omega_{qz})-2\bar\omega_{tz}(2\omega_2+\bar\omega_{2q}).
\label{dDqDz}
\end{gather}
For example, the f\/irst equation in~\eqref{dDqDz} is obtained from the Jacobi identity
$[[\delta,\Delta^q],\Delta^z] + [[\Delta^q,\Delta^z],\delta] + [[\Delta^z,\delta],\Delta^q] = 0$, while the
third and fourth equations are obtained from the single Jacobi identity $[[\delta,\Delta^z],\bar\Delta^z] +
[[\Delta^z,\bar\Delta^z],\delta] + [[\bar\Delta^z,\delta],\Delta^z] = 0$.

Still more resolving equations follow from the def\/initions~\eqref{omtz}--\eqref{ombqz} of the third-order invariants that appear as coef\/f\/icients of commutator algebra of invariant dif\/ferentiations.
Obvious consequences are obtained by invariant cross dif\/ferentiations of each pair of the three equations
in~\eqref{omtz}
\begin{gather}
\label{Dqtt}
\Delta^q(\omega_{tt})=\delta(\omega_{tz}),
\qquad
\bar\Delta^q(\omega_{tt})=\delta(\bar\omega_{tz}),
\qquad
\bar\Delta^q(\omega_{tz})=\Delta^q(\bar\omega_{tz}).
\end{gather}
Invariant cross dif\/ferentiation of the f\/irst equation in~\eqref{omtz} and~\eqref{omqz},~\eqref{ombqz}
written in the form
\begin{gather}
\label{Dzvt}
\Delta^z(v_{tt})=-2\omega_{qz}-2qv_{tq},
\qquad
\bar\Delta^z(v_{tt})=-2\bar\omega_{qz}-2\bar qv_{t\bar q}
\end{gather}
yields
\begin{gather}
\label{Dztt}
\Delta^z(\omega_{tt})=-2\delta(\omega_{qz})+\omega_{tz}^2-2\omega_{tz},
\qquad
\bar\Delta^z(\omega_{tt})=-2\delta(\bar\omega_{qz})+\bar\omega_{tz}^2-2\bar\omega_{tz}.
\end{gather}
Invariant cross dif\/ferentiations of the second and third equations in~\eqref{omtz} together
with~\eqref{omqz} and~\eqref{ombqz}, respectively, set in the form $\delta(qv_{tq}) = \omega_{tz}$ and
$\Delta^q(qv_{tq}) = \omega_{qz} + 2qv_{tq}$, together with complex conjugate equations reproduce f\/irst
two equations in~\eqref{dDqDz}.

Invariant cross dif\/ferentiations of the second and third equations in~\eqref{omtz} together with
equations~\eqref{omqz} and~\eqref{ombqz}, respectively, taken in the form~\eqref{Dzvt}, yield
\begin{gather}
\label{Dztz}
\Delta^z(\omega_{tz})=-2\Delta^q(\omega_{qz})+\omega_{qz}(\omega_{tz}+2),
\qquad
\bar\Delta^z(\bar\omega_{tz})=-2\bar\Delta^q(\bar\omega_{qz})+\bar\omega_{qz}(\bar\omega_{tz}+2).
\end{gather}

Invariant cross dif\/ferentiations of the second equation in~\eqref{omtz} and equation~\eqref{ombqz}, taken
in the form~\eqref{Dzvt}, and also of the third equation in~\eqref{omtz} together with~\eqref{omqz} in the
form~\eqref{Dzvt} yield
\begin{gather}
\bar\Delta^z(\omega_{tz})=-2\Delta^q(\bar\omega_{qz})-4\omega_{2t}-\omega_2\omega_{tt}\omega_{tz},
\nonumber
\\
\Delta^z(\bar\omega_{tz})=-2\bar\Delta^q(\omega_{qz})-4\omega_{2t}-\omega_2\omega_{tt}\bar\omega_{tz}.
\label{Dzomtz}
\end{gather}
Finally, the invariant cross dif\/ferentiation of equations~\eqref{omqz} and~\eqref{ombqz} taken in the
form~\eqref{Dzvt} yields the last resolving equation
\begin{gather}
\label{Dzqz}
\bar\Delta^z(\omega_{qz})=\Delta^z(\bar\omega_{qz})+2\omega_2(\omega_{tz}-\bar\omega_{tz}).
\end{gather}
Thus, the complete set of resolving equations consists of
equations~\eqref{DFG},~\eqref{Dom},~\eqref{dDqDz},~\eqref{Dqtt}, \eqref{Dztt},~\eqref{Dztz},~\eqref{Dzomtz}
and~\eqref{Dzqz}.

\section{Some solutions of the extended system}
\label{sec-sol}

In this section we replace the task of solving the set of resolving equations by a~simpler problem of
f\/inding particular solutions for the commutator algebra of operators of invariant
dif\/ferentiation~\cite{Martina/Sheftel/Winternitz,Nutku/Sheftel2001,Sheftel2002}.
As a~result, we f\/ind some solutions of the Boyer--Finley equation which look very nontrivial and seem to
be new.
They also satisfy all other equations~\eqref{e1}--\eqref{e4} of the extended system and therefore they
also solve the Legendre-transformed CMA equation~\eqref{8a}.
In the next section, we obtain a~noninvariant solution of CMA by applying an inverse Legendre
transformation to one of the obtained solutions.

The extended system~\eqref{e1}--\eqref{e4} contains, besides Boyer--Finley equation~\eqref{e1} (from now
on denoted as BF), the transformed Monge--Amp\`ere equation which is a~consequence of this system.
Therefore, solutions of this system satisfy simultaneously BF and the transformed CMA and hence provide
a~lift of solutions of~\eqref{e1} to those of CMA.

With our choice of the symmetry for the group foliation, non-invariant solutions of the Boyer--Finley
equation, obtained in~\cite{Calderbank/Tod,Martina/Sheftel/Winternitz} and used in~\cite{Nutku/Sheftel2013}
to generate heavenly metrics, cannot be lifted to noninvariant solutions of CMA.

Therefore, in order to solve the resolving equations of the group foliation, constructed for the extended
system, we have to discover some other solutions of the BF equation~\eqref{e1}, which by construction will
be compatible with all other equations of our system~\eqref{e1}--\eqref{e4}.
However, we have too many resolving equations to solve.
Because of that, we use instead the strategy applied in our paper~\cite{Martina/Sheftel/Winternitz} to the
single BF equation, namely, to consider the commutator algebra~\eqref{dD} of operators of invariant
dif\/ferentiation and make some ansatz simplifying this algebra.
The most obvious ansatz is to make as many commutators as possible to vanish
\begin{gather}
\omega_{tt}=v_{ttt}=0,
\qquad
\omega_{tz}=0\ \Rightarrow\ v_{ttq}=0,
\qquad
\bar\omega_{tz}=0\ \Rightarrow\ v_{tt\bar q}=0,
\label{om_ttz}
\\
\omega_{qz}=0\ \Rightarrow\ qv_{ttz}+v_{tq}=0,
\qquad
\bar\omega_{qz}=0\ \Rightarrow\ \bar qv_{tt\bar z}+v_{t\bar q}=0,
\label{om_qz}
\end{gather}
so that the only nonzero commutators are $[\Delta^q,\Delta^z] = 2\Delta^z$,
$[\bar\Delta^q,\bar\Delta^z] = 2\bar\Delta^z$.
The f\/irst equation in~\eqref{om_ttz} is very restrictive since it admits only quadratic $t$-dependence
\begin{gather}
\label{quadr}
v=\frac{\alpha}{2}t^2+\beta t+\gamma,
\end{gather}
where $\alpha$, $\beta$, $\gamma$ are functions of $q$, $\bar q$, $z$, $\bar z$.
Plugging ansatz~\eqref{quadr} in other equations of the extended system, we end up with the following
solution to all of six equations of this system
\begin{gather}
v=-\frac{t^2}{2}\big[\ln{\kappa'(z)}+\ln{\bar\kappa'(\bar z)}\big]+t\left[\sigma(z)+q^2\frac{\kappa''(z)}
{2\kappa'(z)}+\bar\sigma(\bar z)+\bar q^2\frac{\bar\kappa''(\bar z)}{2\bar\kappa'(\bar z)}\right]
\nonumber
\\
\phantom{v=}{}+2q\bar q\sqrt{\kappa'(z)\bar\kappa'(\bar z)}
-\kappa(z)\bar\kappa(\bar z)+\frac{q^4(3\kappa^{\prime\prime\,2}-2\kappa'\kappa''')}{48\kappa'^2}
+\frac{\bar q^4(3\bar\kappa^{\prime\prime\,2}-2\bar\kappa'\bar\kappa''')}{48\bar\kappa'^2}
\nonumber
\\
\phantom{v=}{} -\frac{q^2}{2}\sigma'(z)-\frac{\bar q^2}{2}
\bar\sigma'(\bar z)+q\nu(z)+\bar q\bar\nu(\bar z)+\rho(z)+\bar\rho(\bar z).
\label{zerocom}
\end{gather}

To get more general dependence on $t$, we skip the f\/irst ansatz in~\eqref{om_ttz} and also the
ansatz~\eqref{om_qz}, since the latter does not imply the vanishing of the commutators
$[\Delta^q,\Delta^z]$ and its complex conjugate.
Thus, we now make the only ansatz $v_{ttq} = 0$ and $v_{tt\bar q} = 0$.

Then, without making any more assumptions, we obtain the two following solutions to the whole extended
system:
\begin{gather}
v=-(t+C)^2\left[\ln{(t+C)}-\frac{1}{2}\right]-t^2\left[\frac{1}{2}\ln{(c_1+\bar c_1)}-\ln{a}-1\right]
\nonumber
\\
\hphantom{v=}{}+t\left[2C\ln{a}-\frac{(d-\bar d)^2}{4a}+\varphi_0+\bar\varphi_0\right]+C^2\ln{a}
-\frac{C(d-\bar d)^2}{4a}+\psi_0+\bar\psi_0 
\nonumber
\\
\hphantom{v=}{} +(t+C)\left\{\frac{2\sqrt{c_1\bar c_1}q\bar q}{a}-q^2\frac{c_1}{a}+q\left[\frac{d'}{\sqrt{c_1}}
-\frac{\sqrt{c_1}(d-\bar d)}{a}\right]-\bar q^2\frac{\bar c_1}{a}\right. 
\nonumber
\\
\left.
\hphantom{v=}{} +\bar q\left[\frac{\bar d'}{\sqrt{\bar c_1}}-\frac{\sqrt{\bar c_1}(d-\bar d)}{a}\right]\right\}
-\frac{q^3}{6}\frac{d''}{\sqrt{c_1}}+\frac{q^2}{2}\left(\frac{d^{\prime\,2}}{4c_1}-\frac{2Cc_1}{a}
-\varphi'_0\right)
\nonumber
\\
\hphantom{v=}{}
-\frac{\bar q^3}{6}\frac{\bar d''}{\sqrt{\bar c_1}}+\frac{\bar q^2}{2}\left(\frac{\bar d^{\prime\,2}}
{4\bar c_1}-\frac{2C\bar c_1}{a}-\bar\varphi'_0\right)+q\rho_1+\bar q\bar\rho_1 ,
\label{C}
\end{gather}
where $C\ne 0$ is a~real constant, $a = \bar a = \bar c_1z + \bar c_1\bar z + c_0$ with constant $c_1$, $\bar
c_1$, $c_0$, $d = d(z)$, $\bar d = \bar d(\bar z)$, $\rho_1 = \rho_1(z)$, $\bar\rho_1 = \bar\rho_1(\bar z)$,
$\varphi_0 = \varphi_0(z)$, $\bar\varphi_0 = \bar\varphi_0(\bar z)$, $\psi_0 = \psi_0(z)$, $\bar\psi_0 =
\bar\psi_0(\bar z)$ are arbitrary functions and the primes denote derivatives of functions of a~single
variable.

The second solution obtained with the same ansatz has the form
\begin{gather}
v=-t^2\left[\ln{t}+\frac{1}{2}(\ln{a'}+\ln{\bar a'})-\ln{(a+\bar a)}-\frac{3}{2}\right]
\nonumber
\\
\hphantom{v=}{}+t\left\{2q\bar q\frac{\sqrt{a'\bar a'}}{a+\bar a}+q^2\left(\frac{a''}{2a'}-\frac{a'}{a+\bar a}
\right)+\bar q^2\left(\frac{\bar a''}{2\bar a'}-\frac{\bar a'}{a+\bar a}\right)\right. 
\nonumber
\\
\left.
\hphantom{v=}{} +q\left[\frac{d'}{\sqrt{a'}}-\frac{\sqrt{a'}(d-\bar d)}{a+\bar a}\right]+\bar q\left[\frac{\bar d'}
{\sqrt{\bar a'}}+\frac{\sqrt{\bar a'}(d-\bar d)}{a+\bar a}\right]-\frac{(d-\bar d)^2}{4(a+\bar a)}
+\varphi_0+\bar\varphi_0\right\} 
\nonumber
\\
\hphantom{v=}{} +q^4\left(\frac{a^{\prime\prime\,2}}{16a^{\prime\,2}}-\frac{a'''}{24a'}\right)+\frac{q^3}{6}
\left(\frac{a''d'}{a^{\prime\,3/2}}-\frac{d''}{\sqrt{a'}}\right)+q^2\left(\frac{d^{\prime 2}}{8a'}
-\frac{\varphi'_0}{2}\right)+q\rho_1+\psi_0
\nonumber
\\
\hphantom{v=}{}+\bar q^4\left(\frac{\bar a^{\prime\prime\,2}}{16\bar a^{\prime\,2}}-\frac{\bar a'''}{24\bar a'}
\right)+\frac{\bar q^3}{6}\left(\frac{\bar a''\bar d'}{\bar a^{\prime\,3/2}}-\frac{\bar d''}{\sqrt{\bar a'}}
\right)+\bar q^2\left(\frac{\bar d^{\prime\,2}}{8\bar a'}-\frac{\bar\varphi'_0}{2}
\right)+\bar q\bar\rho_1+\bar\psi_0,\label{zeroC}
\end{gather}
where $a=a(z)$, $\bar a=\bar a(\bar z)$, $d=d(z)$, $\bar d=\bar d(\bar z)$, $\rho_1 = \rho_1(z)$, $\bar\rho_1 =
\bar\rho_1(\bar z)$, $\varphi_0 = \varphi_0(z)$, $\bar\varphi_0 = \bar\varphi_0(\bar z)$, $\psi_0 =
\psi_0(z)$, $\bar\psi_0 = \bar\psi_0(\bar z)$ are all arbitrary functions of a~single variable.
It is interesting to note that the functions $a$ and $\bar a$ come from the general solution
\begin{gather*}
\varepsilon(z,\bar z)=\frac{\sqrt{a'(z)\bar a'(\bar z)}}{a+\bar a}
\end{gather*}
of the Liouville equation $(\ln{\varepsilon^2})_{z\bar z} = 2\varepsilon^2$.

We note that in the process we have obtained some solutions~\eqref{zerocom},~\eqref{C}, and~\eqref{zeroC}
of the Boyer--Finley equation in the form~\eqref{e1}, which seem to be new.

We also mention separable solutions of the Boyer--Finley equation, which reduce in essence to solutions of
the Liouville equation, obtained by Tod~\cite{Tod1995} and recently by Dunajski et
al.~\cite{{Dunajski2013}}.
These solutions turn out to be invariant with respect to one-parameter subgroup of the conformal symmetry
group of the Boyer--Finley equation, so it generates a~metric with at least two Killing vectors, one of
which comes from the BF equation, being itself a~rotational symmetry reduction of CMA.
We see that the Liouville equation arises also in the present paper but in a~dif\/ferent context, not
implying separability of our solutions.

\section[Simultaneous solutions to Boyer-Finley and complex Monge-Amp\`ere equations]{Simultaneous solutions to Boyer--Finley \\ and complex Monge--Amp\`ere equations}
\label{sec-CMAsol}

We study here the most nontrivial solution~\eqref{zeroC} of the extended system.
It should be transformed to a~simultaneous solution $u$ of the Legendre-transformed CMA~\eqref{8a} and
Boyer--Finley~\eqref{cmarot} equations together with other equations~\eqref{Ia} and~\eqref{IIa} of the
extended system.
To achieve this, we apply to solution~\eqref{zeroC} one-dimensional Legendre transformation contained in
formula~\eqref{legendre}: $\rho=-w_t,\; u = w - tw_t$ together with $w=v_t$ to obtain
\begin{gather*}
t=\frac{(a+\bar a)}{\sqrt{a'\bar a'}}\,e^{\rho/2}
\end{gather*}
and the solution becomes
\begin{gather}
u=\frac{2(a+\bar a)}{\sqrt{a'\bar a'}} e^{\rho/2}+2q\bar q\frac{\sqrt{a'\bar a'}}{(a+\bar a)}
+q^2\left(\frac{a''}{2a'}-\frac{a'}{a+\bar a}\right)+\bar q^2\left(\frac{\bar a''}{2\bar a'}-\frac{\bar a'}
{a+\bar a}\right)
\nonumber
\\
\hphantom{u=}{}+q\left[\frac{d'}{\sqrt{a'}}-\frac{\sqrt{a'}(d-\bar d)}{a+\bar a}\right]+\bar q\left[\frac{\bar d'}
{\sqrt{\bar a'}}+\frac{\sqrt{\bar a'}(d-\bar d)}{a+\bar a}\right]-\frac{(d-\bar d)^2}{4(a+\bar a)}
+\varphi_0+\bar\varphi_0,
\label{usol}
\end{gather}
where we change the notation of independent variables from $z$, $\bar z$ to $\sigma$, $\bar\sigma$ for the
functions $a$, $\bar a$, $d$, $\bar d$, $\varphi_0$, $\bar\varphi_0$.
The expression~\eqref{usol} is a~solution of the Legendre-transformed CMA~\eqref{8a}.
To transform back~\eqref{usol} to a~solution of the CMA equation~\eqref{modcma}, which after changing the
notation $v\mapsto\Omega$ and setting $s=\sigma e^{\rho/4}$, $\bar s=\bar\sigma e^{\rho/4}$ becomes
\begin{gather}\label{cmapar}
\Omega_{p\bar p}\Omega_{\sigma\bar\sigma}-\Omega_{p\bar\sigma}\Omega_{\sigma\bar p}=e^{\rho/2}
\end{gather}
we apply to solution~\eqref{usol} the inverse Legendre transformation of the form~\eqref{legmod}
\begin{gather}
\label{legtoOM}
\Omega=u-qu_q-\bar qu_{\bar q},
\qquad
p=-u_q,
\qquad
\bar p=-u_{\bar q}.
\end{gather}
To eliminate~$q$, $\bar q$ from the solution, the two latter equations in~\eqref{legtoOM} can easily be solved
for~$q$ and~$\bar q$
\begin{gather*}
q=\alpha p+\beta\bar p+\gamma,
\qquad
\bar q=\bar\alpha\bar p+\beta p+\bar\gamma,
\end{gather*}
where
\begin{gather}
\alpha=\frac{A}{\Delta},
\qquad
\bar\alpha=\frac{\bar A}{\Delta},
\qquad
\beta=\bar\beta=\frac{B}{\Delta},
\qquad
\gamma=\frac{C}{\Delta},
\qquad
\bar\gamma=\frac{\bar C}{\Delta},
\nonumber
\\
 A=\bar a'[2a^{\prime 2}-a''(a+\bar a)],
\qquad
\bar A=a'[2\bar a^{\prime 2}-\bar a''(a+\bar a)],
\label{notat}
\\
B=\bar B=2(a'\bar a')^{3/2},
\qquad
C=\frac{1}{\sqrt{a'}}(d^{\prime}\bar A+a'\bar D),
\qquad
\bar C=\frac{1}{\sqrt{\bar a'}}(\bar d^{\prime}A+\bar a'D),
\nonumber
\\
 D=\bar a'[2a'd'-a''(d-\bar d)],
\qquad
\bar D=a'[2\bar a'\bar d\,'+\bar a''(d-\bar d)],
\nonumber
\\
 \Delta=a''\bar a''(a+\bar a)-2a''\bar a^{\prime 2}-2\bar a''a^{\prime 2}.
\nonumber
\end{gather}
Then according to the formula~\eqref{legtoOM} for $\Omega$, solution~\eqref{usol} f\/inally becomes
\begin{gather}
\Omega=\frac{\bar\alpha}{2}p^2+\frac{\alpha}{2}
\bar p^2+\beta p\bar p+\gamma p+\bar\gamma\bar p+\frac{\Delta(\alpha\gamma^2+\bar\alpha\bar\gamma^2-2\beta\gamma\bar\gamma)}{2a'\bar a'(a+\bar a)}
\nonumber
\\
\hphantom{\Omega=}{}
-\frac{(d-\bar d)^2}{4(a+\bar a)}+\frac{2(a+\bar a)}{\sqrt{a'\bar a'}}e^{\rho/2}
+\varphi_0+\bar\varphi_0.
\label{sol1}
\end{gather}

\section{Anti-self-dual Ricci-f\/lat metric of Euclidean signature}
\label{sec-metric}
\setcounter{equation}{0}

It is well known~\cite{Plebanski} that solutions of the CMA equation~\eqref{modcma} govern the K\"ahler
metric
\begin{gather}
\label{kahler}
ds^2=2(v_{p\bar p}dpd\bar p+v_{p\bar s}dpd\bar s+v_{s\bar p}dsd\bar p+v_{s\bar s}dsd\bar s),
\end{gather}
which is anti-self-dual (ASD) Einstein vacuum (Ricci-f\/lat) metric with Euclidean signature.
The transformation $s=\sigma e^{\rho/4}$, $\bar s=\bar\sigma e^{\rho/4}$, which maps equation~\eqref{modcma}
into equation~\eqref{cmapar}, does not change the metric~\eqref{kahler} and hence the solution~\eqref{sol1}
of the equation~\eqref{cmapar} can be used as a~potential of the K\"ahler metric
\begin{gather}
\label{Ommetr}
ds^2=2(\Omega_{p\bar p}dpd\bar p+\Omega_{p\bar\sigma}dpd\bar\sigma+\Omega_{\sigma\bar p}
d\sigma d\bar p+\Omega_{\sigma\bar\sigma}d\sigma d\bar\sigma).
\end{gather}
Plugging our solution~\eqref{sol1} into the metric~\eqref{Ommetr} we obtain an explicit Einstein vacuum
metric with Euclidean signature
\begin{gather}
ds^2=2\beta dpd\bar p+2(p\bar\alpha_{\bar\sigma}+\bar p\beta_{\bar\sigma}+\gamma_{\bar\sigma}
)dpd\bar\sigma+2(\bar p\alpha_\sigma+p\beta_\sigma+\bar\gamma_\sigma)d\bar pd\sigma
\nonumber
\\
\left.
\hphantom{ds^2=}{}
+\biggl\{p^2\bar\alpha_{\sigma\bar\sigma}+\bar p^2\alpha_{\sigma\bar\sigma}
+2p\bar p\beta_{\sigma\bar\sigma}+2p\gamma_{\sigma\bar\sigma}+2\bar p\bar\gamma_{\sigma\bar\sigma}
\right.\nonumber\\
\left.
\hphantom{ds^2=}{}
+
\left[\frac{\Delta(\alpha\gamma^2+\bar\alpha\bar\gamma^2-2\beta\gamma\bar\gamma)}{a'\bar a'(a+\bar a)}
-\frac{(d-\bar d)^2}{2(a+\bar a)}\right]_{\sigma\bar\sigma}-\frac{2}{\beta}e^{\rho/2}\right\}
d\sigma d\bar\sigma.
\label{metr1}
\end{gather}

From the def\/initions~\eqref{notat} of the metric coef\/f\/icients in~\eqref{metr1}, we see that the only
singularities of the metric in a~bounded domain are zeros of the denominator
\begin{gather}
\label{sing}
\Delta=a''\bar a''(a+\bar a)-2a''\bar a^{\prime 2}-2\bar a''a^{\prime 2}=0.
\end{gather}
Indeed, the existence condition for our solution is $a'(\sigma)\cdot\bar a'(\bar\sigma)\ne 0$, so that $a$
and $\bar a$ are not constant and hence $a(\sigma)+\bar a(\bar\sigma)\ne 0$.
The general solution for the singularity condition~\eqref{sing} in the case of $a''\cdot\bar a''\ne 0$ is
\begin{gather}
\label{singloc}
a=i\lambda-\frac{1}{a_1\sigma+a_0},
\qquad
\bar a=-i\lambda-\frac{1}{\bar a_1\bar\sigma+\bar a_0}
\end{gather}
and if $a''= \bar a''= 0$, we have
\begin{gather}
\label{singlin}
a=a_1\sigma+a_0,
\qquad
\bar a=\bar a_1\bar\sigma+\bar a_0.
\end{gather}
Here $\lambda$ and $a_0$, $a_1$ are real and complex constants, respectively.
Thus, avoiding the choices~\eqref{singloc} and~\eqref{singlin} for~$a$ and~$\bar a$, we have the
metric~\eqref{metr1} free of singularities in a~bounded domain.

To compute Riemann curvature two-forms, we choose the tetrad coframe~\cite{Eguchi/Gilkey/Hanson} to be
\begin{gather*}
e^1=\frac{1}{\Omega_{p\bar p}} (\Omega_{p\bar p}d p+\Omega_{z\bar p}d z),
\qquad
e^2=\Omega_{p\bar p}d\bar p+\Omega_{p\bar z}d\bar z
\qquad
e^3=\frac{e^{\rho/2}}{\Omega_{p\bar p}}\,d z,
\qquad
e^4=d\bar z,
\end{gather*}
so that the metric~\eqref{Ommetr} with the aid of equation~\eqref{cmapar} takes the form $ds^2 = 2\big(e^1e^2 +
e^3e^4\big)$, where we plug in the solution~\eqref{sol1} for~$\Omega$.

Using the computer algebra package EXCALC under REDUCE~\cite{Excalc}, we calculated Riemannian curvature
two-forms~\cite{Eguchi/Gilkey/Hanson} for the metric~\eqref{metr1}
\begin{gather}
R^1_{\;1}=2e^{-\rho/2}\frac{|a'|^5}{\Delta^3}
\Big|2a'''a'-3(a'')^2\Big|^2\big(e^1\wedge e^2-e^3\wedge e^4\big)=-R^2_{\;2}=-R^3_{\;3}=R^4_{\;4},
\nonumber
\\
R^1_{\;2}=R^1_{\;4}=R^2_{\;1}=R^2_{\;3}=R^3_{\;2}=R^3_{\;4}=R^4_{\;1}=R^4_{\;3}=0
\label{curv}
\end{gather}
with the remaining curvature two-forms being too lengthy for presentation here.
For example
\begin{gather}
R^1_{\;3}=\frac{(\bar a')^3e^{-\rho}}{\sqrt{a'}\Delta^3}\big\{a''(4a'''a'-3a''^{2})(5\Delta+18a'^{2}
\bar a'')
\nonumber
\\
\hphantom{R^1_{\;3}=}{}
-12a'^{2}a'''^{2}[(a+\bar a)\bar a''-2\bar a'^{2}]+4a'^{2}a^{\prime\prime\prime\prime}\Delta\big\}e^2\wedge e^3
\nonumber
\\
\hphantom{R^1_{\;3}=}{}
-2e^{-\rho/2} \frac{a'^{2}\sqrt{\bar a'}}{\Delta^3} |2a'a'''-3a''^{2}|^2e^1\wedge e^4.
\label{R_13}
\end{gather}
We see that the only singularities of the curvature in a~bounded domain are poles at $\Delta=0$ together
with $a'\cdot\bar a'=0$ which are the same as those of the metric.
We note that our solution~\eqref{sol1} does not exist if the singularities conditions are satisf\/ied.

We observe that though the metric~\eqref{metr1} contains $p$ and $\bar p$, the curvature is independent of
these variables.
This is due to the quadratic dependence of the metric on $p$ and $\bar p$.
Therefore, the curvature components do not vanish outside of a~bounded domain, so our metric is not of an
instanton type.

One may wonder if there are choices of $a(\sigma)$, $\bar a(\bar\sigma)$ for which all components of the
Riemannian tensor vanish, so that our solution describes a~f\/lat space.
The inspection of our result for the Riemannian curvature two-forms~\eqref{curv},~\eqref{R_13} gives the
conditions for zero curvature to be of the form
\begin{gather*}
a'''=\frac{3a''^{2}}{2a'},
\qquad
\bar a'''=\frac{3\bar a''^{2}}{2\bar a'},
\end{gather*}
which yield
\begin{gather*}
a=-\frac{4}{k(k\sigma+l)},
\qquad
\bar a=-\frac{4}{\bar k(\bar k\bar\sigma+\bar l)},
\end{gather*}
where $k$, $l$, $\bar k$, $\bar l$ are arbitrary constants.
This result coincides with the singularity condition~\eqref{singloc} at $\lambda=0$ for which our
solution~\eqref{sol1} does not exists and therefore this is not allowed, so that we will never have
a~f\/lat space for any allowed choice of $a$, $\bar a$.

We have succeeded in converting the solution~\eqref{usol} into solution~\eqref{sol1} of CMA
equation~\eqref{cmapar} because of the quadratic dependence of solution~\eqref{usol} on $q$, $\bar q$, which
allowed us to eliminate these variables in terms of $p$, $\bar p$ by solving a~linear system.
For a~more general dependence of solutions on $q$, $\bar q$, this transition could happen to be impossible to
be performed explicitly.
In this case, we need to transform K\"ahler metric~\eqref{kahler} to the one that involves a~solution $u$
of the Legendre-transformed CMA~\eqref{8a} (solution~\eqref{usol} in our example) by applying the inverse
Legendre transformation~\eqref{legmod} with~$v$ replaced by $\Omega$ to the metric~\eqref{Ommetr}.
The resulting metric has the form
\begin{gather}
ds^2=\frac{2}{\Delta_-}\big\{u_{q\bar q}^2\big(u_{qq}dq^2+u_{\bar q\bar q}d\bar q^2\big)+\Delta_+u_{q\bar q}
dqd\bar q+u_{qq}u_{\bar qz}^2dz^2+u_{\bar q\bar q}u_{q\bar z}^2d\bar z^2
\nonumber
\\
\hphantom{ds^2=}{}+(\Delta_-u_{z\bar z}+2u_{q\bar q}u_{q\bar z}u_{\bar qz})dzd\bar z+2u_{q\bar q}(u_{qq}
u_{\bar qz}dq dz+u_{\bar q\bar q}u_{q\bar z}d\bar qd\bar z)
\nonumber
\\
\hphantom{ds^2=}{}
+\Delta_+(u_{q\bar z}dqd\bar z+u_{\bar qz}d\bar qdz)\big\},
\label{legmetr}
\end{gather}
where $\Delta_- = u_{qq}u_{\bar q\bar q} - u_{q\bar q}^2$ and $\Delta_+ = u_{qq}u_{\bar q\bar q} + u_{q\bar
q}^2$.
Metric~\eqref{legmetr} is again anti-self-dual and Ricci-f\/lat with Euclidean signature for any solution
$u$ of the transformed CMA equation~\eqref{8a}.

\section{Invariant and noninvariant solutions}
\label{sec-invsol}

We are interested here in ASD Ricci-f\/lat metrics of Euclidean signature that do not admit any Killing
vectors.
This implies solutions of CMA equation for the K\"ahler potential of the metric to be noninvariant
solutions of CMA.
This means that we allow only solution manifolds noninvariant under point symmetries of CMA equation.
For a~set of solutions of the Boyer--Finley equation we presented such an analysis in detail
in~\cite{Martina/Sheftel/Winternitz}.
In our case, for CMA equation~\eqref{cmapar}, dependent of an extra parameter $\rho$, we have the following
generators of one-parameter point symmetries subgroups
\begin{gather}
X_{a_1}=a_1(\rho)(4\partial_\rho+\Omega\partial_\Omega),
\qquad
Y_b=b(\rho)(p\partial_p+\bar p\partial_{\bar p}+\Omega\partial_\Omega),
\nonumber
\\
Z_{c_1}=i c_1(\rho)(\sigma\partial_\sigma-\bar\sigma\partial_{\bar\sigma}),
\qquad
V_g=g_p\partial_\sigma-g_\sigma\partial_p,
\qquad
\bar V_{\bar g}=\bar g_{\bar p}\partial_{\bar\sigma}-\bar g_{\bar\sigma}\partial_{\bar p},
\nonumber
\\
W_h=h\partial_\Omega,
\qquad
\bar W_{\bar h}=\bar h\partial_{\Omega},
\label{symcma}
\end{gather}
where $g=g(p,\sigma,\rho)$, $\bar g=\bar g(\bar p,\bar\sigma,\rho)$ and $h=h(p,\sigma,\rho)$, $\bar h=\bar
h(\bar p,\bar\sigma,\rho)$ are arbitrary holomorphic and anti-holomorphic functions of two complex and one
real variable, $a_1(\rho)$, $b(\rho)$, and $c_1(\rho)$ are arbitrary real-valued functions of a~single
variable.
We present here the table of commutators of the symmetry generators where the value of the commutator of
the generators standing at the $i$th row and at the $j$th column is given at the intersection of the $i$th
row and $j$th column.
The primes denote derivatives of functions of a~single variable $\rho$.
\begin{table}[ht]
\centering
\caption{Commutators of point symmetries of parameter-dependent CMA.\label{commuttab}}
\vspace{1mm}
\begin{tabular}{c|c|c|c|c|c|c|c}
\hline
\tsep{1mm} \bsep{1mm} &$X_{a_1}$&$Y_b$  &$Z_{c_1}$& $V_g$ &$\bar V_{\bar g}$ &$W_h$ &$\bar W_{\bar h}$
\\ \hline
  \tsep{1mm}\bsep{1mm}  $X_{a_1}$ & $0$ & $4Y_{a_1b'}$& $4Z_{a_1c_1'}$ & $4V_{a_1g_\rho}$ &$4\bar V_{a_1\bar g_\rho}$  &$4W_{a_1h_\rho}$   &$4\bar W_{a_1\bar h_\rho}$
\\ \hline
\tsep{1mm}\bsep{1mm}    $Y_b$ &$$& $0$ & $0$   & $V_{b(pg_p-g)}$ &$\bar V_{b(\bar p\bar g_{\bar p}-\bar g)}$&$W_{b(ph_p-h)}$& $\bar W_{b(\bar p\bar h_{\bar p}-\bar h)}$
\\ \hline
\tsep{1mm}\bsep{1mm}    $Z_{c_1}$ &$$  &$$  & $ 0$  &$iV_{c_1(\sigma g_\sigma-g)}$  &$-i\bar V_{c_1(\bar\sigma\bar g_{\bar\sigma}-\bar g)}$& $iW_{c_1\sigma h_\sigma}$& $-i\bar W_{c_1\bar\sigma\bar h_{\bar\sigma}}$
\\ \hline
\tsep{1mm}\bsep{1mm}    $V_g$ &$$& $$& $$  & $0$& $0$& $W_{V_g(h)}$& $0$
\\ \hline
\tsep{1mm}\bsep{1mm} $\bar V_{\bar g}$ &$$& $$& $$& $$& $0$& $0$& $\bar W_{\bar V_{\bar g}(\bar h)}$
\\ \hline
\tsep{1mm}\bsep{1mm} $W_h$ &$$& $$& $$& $$& $$& $0$& $0$
\\ \hline
\tsep{1mm}\bsep{1mm} $\bar W_{\bar h}$ &$$& $$& $$& $$& $$& $$& $0$
\\ \hline
\end{tabular}

\end{table}

For any solution $\Omega=f(p,\bar p,\sigma,\bar\sigma,\rho)$ of equation~\eqref{cmapar},
the condition for this solution to be invariant under a generator $X$ of an arbitrary one-parameter symmetry subgroup of~\eqref{cmapar} has the form
\begin{gather}
\label{invsolcond}
X(f-\Omega)|_{\Omega=f}=0.
\end{gather}
With $X$ equal to a~linear combination of the generators~\eqref{symcma} with arbitrary constant
coef\/f\/icients (in our case absorbed by arbitrary functions in the generators), the invariance
condition~\eqref{invsolcond} for the solution $\Omega = f(p,\bar p,\sigma,\bar\sigma,\rho)$ becomes
\begin{gather}
g_pf_\sigma-g_\sigma f_p+\bar g_{\bar p}f_{\bar\sigma}-\bar g_{\bar\sigma}f_{\bar p}
+b(\rho)(pf_p+\bar pf_{\bar p})+i\tilde c(\rho)(\sigma f_\sigma-\bar\sigma f_{\bar\sigma})
\nonumber
\\
\qquad{}+4\tilde a(\rho)f_\rho-(\tilde a(\rho)+b(\rho))f-h-\bar h=0.
\label{invsol}
\end{gather}
To check the non-invariance of the solution~\eqref{sol1}, one should plug $f$ equal to the right-hand side
of~\eqref{sol1} in the condition~\eqref{invsol} and determine either a~contradiction or some special forms
of the coef\/f\/icient functions in~\eqref{sol1} which should be avoided for a~noninvariant solution.

To simplify the invariance condition, we have to use optimal Lie subalgebras instead of the general
one-dimensional subalgebra used in~\eqref{invsol} with the generator
\begin{gather}
\label{gen1}
X=X_{\tilde{a}(\rho)}+Y_{b(\rho)}+Z_{\tilde c(\rho)}+V_g+\bar V_{\bar g}+W_h+\bar W_{\bar h}.
\end{gather}
For this purpose, we study the adjoint group actions on one-dimensional Lie subalgebras~\cite{Olver}.
Here we must distinguish two cases.

\textbf{Case I:}
$\tilde a(\rho)\ne 0$.
In this case, we use the adjoint group actions
\begin{gather*}
\textrm{Ad} \left(\exp\left(-\frac{1}{4}Y_{e^{\int d\rho/\tilde a}}\right)\right)(X_{\tilde a}
)=X_{\tilde a}-Y_b,
\qquad
\textrm{Ad} \left(\exp\left(-\frac{1}{4}Z_{e^{\int (\tilde c/\tilde a)d\rho}}\right)\right)(X_{\tilde a}
)=X_{\tilde a}-Z_{\tilde c}
\end{gather*}
to eliminate $Y_b$ and $Z_{\tilde c}$ in~\eqref{gen1}, so that the optimal subalgebra becomes
\begin{gather*}
X=X_{\tilde{a}}+V_g+\bar V_{\bar g}+W_h+\bar W_{\bar h}
\end{gather*}
which results in setting $b = \tilde c = 0$ in the invariance condition~\eqref{invsol}:
\begin{gather}
g_pf_\sigma-g_\sigma f_p+\bar g_{\bar p}f_{\bar\sigma}-\bar g_{\bar\sigma}f_{\bar p}
+\tilde a(\rho)(4f_\rho-f)-h-\bar h=0.
\label{invsol1}
\end{gather}

\textbf{Case II:}
$\tilde a(\rho) = 0$.
Then using the adjoint group actions
\begin{gather*}
\textrm{Ad}\!\left(\exp(\varepsilon W_h\right)(Y_b)=Y_b+\varepsilon W_{b(ph_p-h)},
\qquad
\textrm{Ad}\!\left(\exp(\varepsilon V_g\right)(Y_b)=Y_b+\varepsilon V_{\tilde g}
\end{gather*}
together with their complex conjugates, we can eliminate $V_g$, $W_h$, $\bar V_{\bar g}$, $\bar W_{\bar h}$
from~\eqref{gen1} at $\tilde a(\rho) = 0$ and the second optimal one-dimensional subalgebra becomes
\begin{gather}
\label{opt2}
X=Y_b+Z_{\tilde c}.
\end{gather}

The invariance condition~\eqref{invsol} in the Case~II due to the result~\eqref{opt2} for the optimal
subalgebra implies $g=\bar g=\tilde a=h=\bar h = 0$, so that the invariance condition~\eqref{invsol} becomes
\begin{gather}
\label{inv2}
b(\rho)(pf_p+\bar pf_{\bar p}-f)+i\tilde c(\rho)(\sigma f_\sigma-\bar\sigma f_{\bar\sigma})=0.
\end{gather}

After some routine computations we discover that in both Cases~I and~II our solution~\eqref{sol1}
generically does not satisfy the invariance conditions~\eqref{invsol1} and~\eqref{inv2}, respectively, and
hence it is noninvariant in the generic case, that is, with no restrictions on arbitrary functions of one
variable in solution~\eqref{sol1}.
A full classif\/ication of particular choices of the functional parameters that correspond to invariant
solutions presents a~dif\/f\/icult problem which is still expecting its solution.

\section{Conclusion}

The problem of obtaining explicitly the metric of $K3$ gravitational instanton or at least some pieces of
it, which will not admit any Killing vectors (no continuous symmetries), has motivated our search for
non-invariant solutions to the elliptic complex Monge--Amp\`ere equation.
In recent years we have developed three approaches to the latter problem: partner symmetries, that is,
invariance with respect to a~certain nonlocal symmetry, symmetry reduction with respect to symmetry group
parameters introduced explicitly in the theory as new independent variables and a~version of the group
foliation method, which is based on solving commutator algebra relations for operators of invariant
dif\/ferentiation.
In this paper, we have combined all these approaches by introducing explicitly symmetry group parameters
into the extended system of six PDEs, which determine partner symmetries of CMA, performing symmetry
reductions of these equations with respect to the group parameters and, f\/inally, applying the group
foliation to the reduced system.
Since the f\/inal reduced system contains the Boyer--Finley equation together with CMA, though not in the
same variables, a~solution to the extended system provides a~lift from some solutions of the elliptic BF
equation to noninvariant solutions of CMA, that is, from rotationally invariant to noninvariant solutions
of CMA.

To provide an example of our solution procedure, we have chosen the most obvious ansatzes simplifying the
commutator algebra of operators of invariant dif\/ferentiation and obtained some solutions to our extended
system which, after Legendre transformations, became new simultaneous solutions to a~parameter-dependent
CMA equation and the BF equation.
Using the most general of the obtained solutions, we obtained an anti-self-dual Ricci-f\/lat
Einstein--K\"ahler metric with Euclidean signature and computed Riemannian curvature two-forms.
The only singularities of the metric and the curvature, located in a~bounded domain, exist only for a~very
special choice of arbitrary functions of one variable in our solution and therefore they can easily be
avoided.
Considering in detail the conditions for our solution to be invariant under optimal symmetry subgroups of
CMA, we have proved that this is a~noninvariant solution in the generic case (that is, with no special
restrictions on functional parameters) and hence our metric does not admit any Killing vectors.

Our main goal here was to demonstrate how our methods may yield ASD Ricci-f\/lat metrics without Killing
vectors, for which purpose we have chosen simplest possible non-invariant solutions of CMA.
Therefore, it is not surprising that our ansatz for the solution was too restrictive to obtain an instanton
metric, so that the curvature is not concentrated in a~bounded domain.
Even though noninvariant instanton solutions were not found, we believe that the groundwork for future
research has been laid, so that a~more systematic study of possible solutions and also dif\/ferent
reductions of the extended system in group parameters will provide gravitational instanton metrics of
Euclidean signature without Killing vectors.

\subsection*{Acknowledgement}

We thank our referees for their encouragement and criticism which hopefully improved our paper.
The research of M.B.~Sheftel was supported in part by the research grant from Bo\u{g}azi\c{c}i University
Scientif\/ic Research Fund (BAP), research project No.~6324.

\pdfbookmark[1]{References}{ref}
\LastPageEnding

\end{document}